\documentclass[aps,prl,twocolumn,superscriptaddress]{revtex4-2}
\usepackage{graphicx}
\usepackage{float}
\usepackage{booktabs}
\usepackage[caption=false]{subfig}
\usepackage{color}
\usepackage[colorlinks,bookmarks=false,citecolor=blue, urlcolor=blue]{hyperref}
\usepackage{amsmath,changes}

\begin{document}


\title{
Divergent Thermal Expansion and Gr\"uneisen Ratio in a Quadrupolar Kondo Metal}


\author{A. W\"orl}
\email[]{andreas.woerl@physik.uni-augsburg.de}
\affiliation{Experimental Physics VI, Center for Electronic Correlations and Magnetism, University of Augsburg, 86159 Augsburg, Germany}

\author{M. Garst}
\affiliation{Institute for Theoretical Solid State Physics, Karlsruhe Institute of Technology, 76131 Karlsruhe, Germany}
\affiliation{Institute for Quantum Materials and Technology, Karlsruhe Institute of Technology, 76131 Karlsruhe, Germany}

\author{Y. Yamane}
\altaffiliation{Current affiliation: Department of Material Science, Graduate School of Science, University of Hyogo, Kamigori, Hyogo 678-1297, Japan}
\affiliation{Department of Quantum Matter, Graduate School of Advanced Science and Engineering, Hiroshima University, Higashi-Hiroshima 739-8530, Japan}

\author{S. Bachus}
\affiliation{Experimental Physics VI, Center for Electronic Correlations and Magnetism, University of Augsburg, 86159 Augsburg, Germany}

\author{T. Onimaru}
\affiliation{Department of Quantum Matter, Graduate School of Advanced Science and Engineering, Hiroshima University, Higashi-Hiroshima 739-8530, Japan}

\author{P. Gegenwart}
\email[]{philipp.gegenwart@physik.uni-augsburg.de}
\affiliation{Experimental Physics VI, Center for Electronic Correlations and Magnetism, University of Augsburg, 86159 Augsburg, Germany}
\date{\today}

\begin{abstract}
We report on the low-temperature thermal expansion and magnetostriction of the single-impurity quadrupolar Kondo candidate Y$_{1-x}$Pr$_{x}$Ir$_2$Zn$_{20}$. In the dilute limit, we find a quadrupolar strain that possesses a singular dependence on temperature $T$, 
$\varepsilon_{\mathrm{u}} \sim H^2 \log 1/T$,  for a small but finite magnetic field $H$. Together with the previously reported anomalous specific heat $C$, this implies a quadrupolar Gr\"uneisen ratio  $\Gamma_u = \partial_T \varepsilon_{\mathrm{u}}/C \sim  H^2/(T^2 \log 1/T)$ whose divergence for finite $H$ is consistent with the scenario of a quadrupolar Kondo effect. In addition, we find a singular behavior of the isotropic strain $\varepsilon_{\mathrm{B}}$ in zero  magnetic field resulting in a divergence of  both the volume thermal expansion and the volume Gr\"uneisen parameter. We speculate that this behavior might be also induced by putative Kondo correlations via elastic anharmonicities or static strain disorder. 
\end{abstract}

\maketitle

In solid state physics, non-Fermi liquid phases describe unconventional metallic states of matter. Such exotic phases have been extensively studied in the framework of heavy fermion (HF) quantum criticality \cite{Gegenwart08}. A magnetic quantum critical point typically forms in Ce- and Yb-based intermetallic systems that are located in between a Kondo screened Fermi-liquid and a long range antiferromagnetically ordered state. The divergence of the Grüneisen parameter, defined as the ratio of volume thermal expansion to specific heat, is a universal signature of pressure sensitive quantum critical points~\cite{ZhuGarst03, Gegenwart16}. 

In addition, there has been a particular focus on exploring non-Fermi liquid states that are related to electric quadrupole moments and on identifying potential links between these states and magnetic HF quantum criticality \cite{OnimaruIzawa16, YamaneOnimaru18, MatsubayashiTanaka12, ShimuraTsujimoto15}.  Quadrupolar ground states form, for instance, in non-Kramers Pr- and U-based intermetallics, given that certain symmetry constraints are fulfilled. Possible cause for unconventional metallic behavior in these materials is the quadrupolar Kondo effect, which was postulated by Cox in 1987 \cite{Cox87}. Here, the simultaneous overscreening of a localized quadrupole moment by two-channels of conduction electrons, which are related to their spin degree of freedom, leads to non-Fermi liquid behavior in specific heat ($C/T\propto \log 1/T$), electrical resisitivity ($\rho/\rho_{0}\propto 1+A\sqrt{T}$) and quadrupole susceptibility ($\chi_{\mathrm{Q}}\propto \log{1/T}$) as well as an unconventional residual entropy of $S= R \log \sqrt{2}$ \cite{CoxZawadowski98}.

Cubic Pr-based 1-2-20 systems are prototypical to study such novel quadrupolar ~\cite{OnimaruKusunose16, SakaiNakatsuji11} as well as higher multipolar related correlation effects~\cite{TsujimotoPRL14, PatriNCom19, PatriPRX20}. PrIr$_2$Zn$_{20}$, for instance, has a well defined quadrupolar non-Kramers $\Gamma_3$ ground state doublet (point group $T_{d}$) \cite{IwasaHiroki13} and displays clear signatures of the quadrupolar Kondo lattice effect \cite{OnimaruIzawa16}, which are cut off by antiferroquadrupolar order at 0.11\,K~\cite{OnimaruMatsumoto11}. Recent studies on highly diluted Y$_{1-x}$Pr$_{x}$Ir$_{2}$Zn$_{20}$ provided direct evidence of the single-impurity quadrupolar Kondo effect, based on characteristic non-Fermi liquid behaviors found in the specific heat, electrical resistivity and elastic constant \cite{YamaneOnimaru18, YanagisawaHidaka19}. Nevertheless, evidence for the residual entropy $S= R \log \sqrt{2}$ remains elusive so far \cite{YamaneOnimaru18}.

The concept of the Gr\"uneisen parameter can be generalized to all irreducible representations of elastic strains. 
Such a generalized Gr\"uneisen parameter is expected to diverge close to a quantum critical point provided that the associated stress couples to a relevant operator of the critical fixed-point \cite{ZhuGarst03,Zacharias2015}. For the single-impurity quadrupolar Kondo fixed-point, this is the case for the quadrupolar stress $\sigma_{\mathrm{u}}$, that breaks the cubic symmetry, and, as a consequence, destabilizes the non-Fermi liquid physics and quenches the residual entropy. This implies that the quadrupolar Gr\"uneisen ratio $\Gamma_{\mathrm{u}} = (\partial_T \varepsilon_{\mathrm{u}})/C$ diverges in a characteristic manner. Here, $\varepsilon_{\mathrm{u}} = (2 \varepsilon_{zz} - \varepsilon_{xx} - \varepsilon_{yy})/\sqrt{3}$ is the strain component of the $\Gamma_3$ doublet that is conjugate to $\sigma_{\mathrm{u}}$ with the corresponding thermal expansion $\alpha_{\mathrm{u}}=\partial_T \varepsilon_{\mathrm{u}}$ and $C = C_{\mathrm{m}}/V_{\mathrm{m}}$ the specific heat with the molar volume $V_\mathrm{m}$ and the molar specific heat $C_\mathrm{m}$. Using a thermodynamic identity, $\Gamma_{\mathrm{u}}= 1/T(\mathrm{d}T/\mathrm{d}\sigma_{\mathrm{u}})_S $ also quantifies the adiabatic change of temperature upon the variation of $\sigma_{\mathrm{u}}$ and thus describes an elastocaloric effect~\cite{Ikeda19}. 

The application of a magnetic field $\boldsymbol H$ will split the non-Kramers $\Gamma_3$ doublet resulting in a finite quadrupole moment $\langle Q \rangle \sim \chi_{\mathrm{Q}} H^2$ of each Pr$^{3+}$ ion, that is proportional to $H^2$ for small fields due to time-reversal symmetry. This induces a local quadrupolar strain via the elastic coupling $g_{\Gamma_3}$, and after averaging over Pr disorder results in a homogeneous strain $\varepsilon_{\mathrm{u}} \sim g_{\Gamma_3} n_{\mathrm{Pr}} \chi_{\mathrm{Q}} H^2/c_{\mathrm{u}}$ where $n_{\mathrm{Pr}}$ is the Pr$^{3+}$ density and $c_{\mathrm{u}} = (c_{11} - c_{12})/2$ the corresponding elastic constant. The quadrupolar Kondo effect predicts $\chi_{\mathrm{Q}} \sim \log 1/T$ resulting in a singular temperature dependence of the quadrupolar strain $\varepsilon_{\mathrm{u}} \sim n_{\mathrm{Pr}} H^2 \log 1/T$. Finally, using that the specific heat at low temperatures is dominated by the contribution of the Pr$^{3+}$ ions, $C_{\mathrm{m}} \sim n_{\mathrm{Pr}} T \log 1/T$ for small magnetic fields \cite{YamaneOnimaru18}, one expects for the  quadrupolar Gr\"uneisen parameter $\Gamma_{\mathrm{u}} \sim  H^2/(T^2 \log{1/T})$. Consequently, thermal expansion and magnetostriction experiments are ideally suited for the investigation of the non-Fermi liquid behavior associated with the quadrupolar Kondo effect.

In this work, we perform such measurements on highly diluted single crystalline Y$_{1-x}$Pr$_x$Ir$_2$Zn$_{20}$. As our key finding, we confirm experimentally that the temperature dependence for the quadrupolar strain $\varepsilon_{\mathrm{u}}$ and the associated quadrupolar Gr\"uneisen parameter $\Gamma_{\mathrm{u}}$ is indeed consistent with the suggested quadrupolar Kondo scenario~\cite{YamaneOnimaru18, YanagisawaHidaka19}. Remarkably, we also find that the volume strain $\varepsilon_{\mathrm{B}}$ shows a similar singular behavior in zero magnetic field although isotropic stress does not directly couple to a relevant operator of the quadrupolar Kondo fixed-point. We speculate that an indirect coupling might be generated either by elastic anharmonicities or by static strain disorder accounting for the experimentally observed divergence in the volume thermal expansion.
 
Ultrahigh-resolution thermal expansion and magnetostriction measurements were carried out in a dilution refrigerator using a miniaturized capacitive dilatometer~\cite{KuechlerWoerl17}. Central to this study is a highly diluted Y$_{1-x}$Pr$_{x}$Ir$_{2}$Zn$_{20}$ single crystal with $x=0.036$, for which relative length changes were measured along a cubic $\langle 100\rangle$ direction, with magnetic fields applied either parallel or perpendicular to the measurement direction. The magnetic field direction is defined as [001] in the following, so that $\varepsilon_{zz}$ and $\varepsilon_{xx}$ correspond to longitudinal $\varepsilon_{\parallel}$ and transverse strain $\varepsilon_{\perp}$, respectively. As the applied field does not break the symmetry within the $(x,y)$ plane, we can assume that $\varepsilon_{xx} = \varepsilon_{yy}$. The measurement of longitudinal and transverse strain in magnetic field then allows to infer the values of the isotropic strain $\varepsilon_{\mathrm{B}} = \varepsilon_{xx} + \varepsilon_{yy} + \varepsilon_{zz}$ and
the quadrupolar strain $\varepsilon_{\mathrm{u}}$, as detailed in the Supplemental Material (SM)~\cite{SupplementalMaterial}\nocite{YamaneOnimaru18PhysicaB, YamaneOnimaru18AIP, PottSchefzyk83, WoerlOnimaru19, LeaLeask62}. An unavoidable side effect of the experimental technique is a small force of approximately $4$\,N \cite{KuechlerWoerl17} acting on the sample along the measurement direction, which corresponds to a tiny unixial stress of a few MPa. To clarify whether this effect has an impact on the deduced relative length changes, we performed complementary measurements on a $[111]$ oriented single crystal with a comparable Pr-concentration of $x=0.033$. For further characterization, a special uniaxial stress capacitive dilatometer \cite{KuechlerStingl16}, that exerts a roughly fifteen times larger force on the sample than the miniaturized dilatometer, was employed. For details on the single crystalline samples examined in this study, see SM \cite{SupplementalMaterial}.

First, we discuss the quadrupolar thermal expansion coefficient $\alpha_{\mathrm{u}} = \partial_T \varepsilon_{\mathrm{u}}$ of Y$_{1-x}$Pr$_{x}$Ir$_{2}$Zn$_{20}$ with $x=0.036$. The respective data is derived from the measurement of the longitudinal $\alpha_{\parallel}$ and the volume thermal expansion~$\beta$ for $\boldsymbol{H}\parallel[001]$ via the the relation $\alpha_{\mathrm{u}}=\sqrt{3}(\alpha_{\parallel}-\beta/3)$. A detailed derivation of this relation and an overview of the data of $\alpha_{\parallel}$ and $\beta$ is provided in the SM~\cite{SupplementalMaterial}.
\begin{figure}
\includegraphics[width=0.48\textwidth]{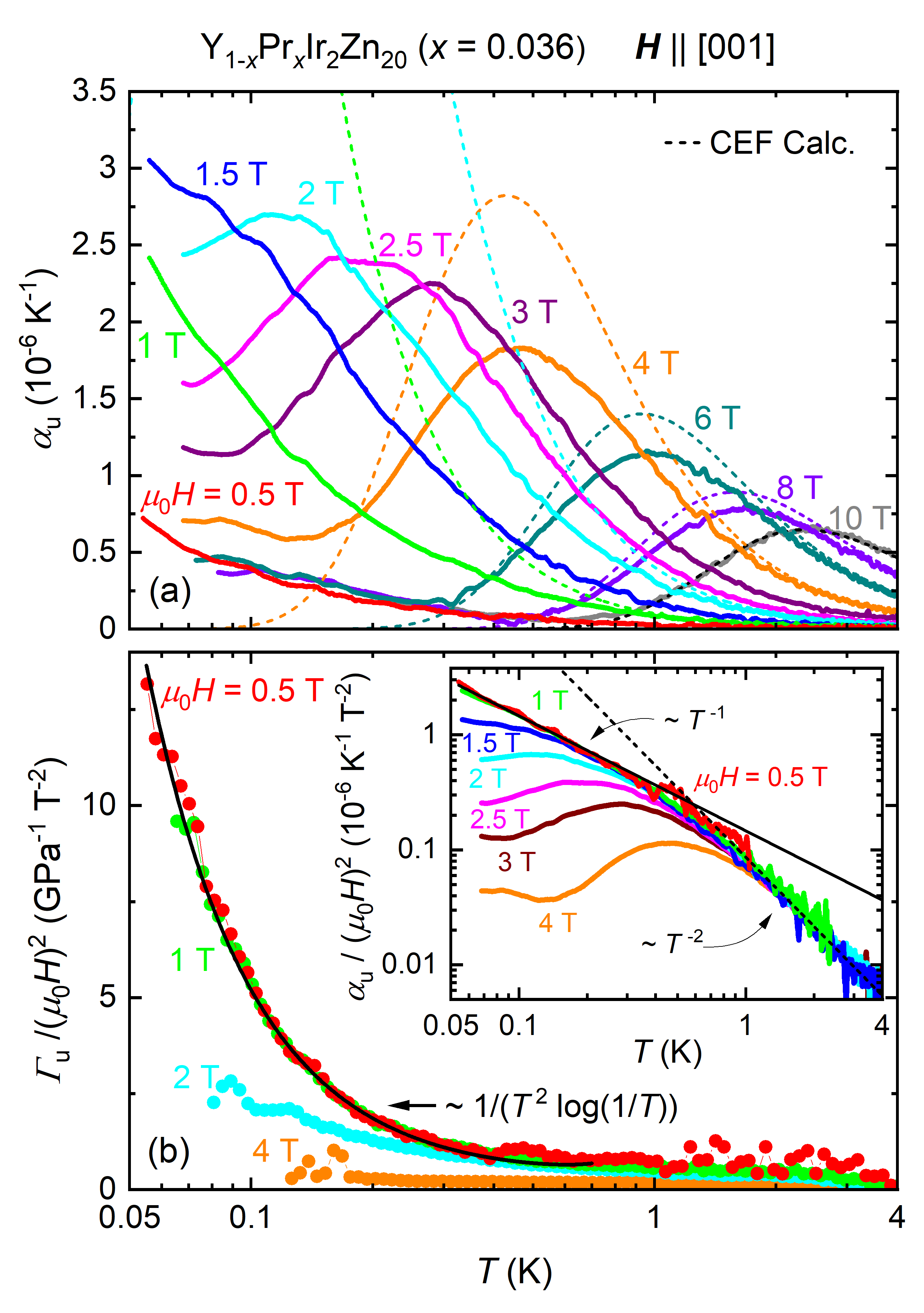}
\caption{(a) Temperature dependence of the quadrupolar thermal expansion coefficient $\alpha_{\mathrm{u}}$ at various magnetic fields $\boldsymbol H\parallel[001]$. The dashed lines are crystal electric field (CEF) calculations. (b) Quadrupolar Grüneisen parameter normalized to magnetic field $\Gamma_{\mathrm{u}}/H^2$ as a function of temperature for $\boldsymbol H\parallel[001]$. The black solid line denotes the theoretically expected temperature dependence of $\Gamma_{\mathrm{u}}$ for the quadrupolar Kondo effect. The inset shows $\alpha_{\mathrm{u}}/H^2$ vs. $T$ on a log-log scale, together with power law divergences as dashed and solid lines.}
\label{fig:fig1}
\end{figure}
The quadrupolar thermal expansion coefficient $\alpha_{\mathrm{u}}$ is shown in Fig.~\ref{fig:fig1}(a) on a logarithmic temperature scale ranging from 0.05\,K to 4\,K for various magnetic fields up to 10\,T. 
For small fields, $\alpha_{\mathrm{u}}$ increases down to lowest temperatures. At around 1.5\,T a maximum develops whose position shifts to higher temperatures as a function of increasing field. At the same time, the height of the maximum continuously decreases with $H$. For $\mu_{0}H\gtrsim 8$ T, this peak is quantitatively captured by crystal electric field (CEF) calculations \cite{SupplementalMaterial} as indicated by the dashed lines.

The ratio $\alpha_{\mathrm{u}}/(\mu_0 H)^2$ is shown in the inset of Fig.~\ref{fig:fig1}(b). 
For temperatures $k_{\mathrm{B}} T \gg \mu_{\mathrm{B}} \mu_0 H$, the data collapses onto a single curve that exhibits a crossover from a $\alpha_{\mathrm{u}}/H^2 \sim \partial_T \chi_{\mathrm{Q}} \sim 1/T$ behavior at low temperature, consistent with the quadrupolar Kondo effect, 
to a $1/T^2$ dependence at high temperature as expected for a Curie susceptibility $\chi_{\mathrm{Q}} \sim 1/T$ of a fully localized $\Gamma_3$ doublet\cite{SupplementalMaterial}. The crossover temperature of $\sim 0.6$\,K is consistent with previous studies \cite{YamaneOnimaru18,YanagisawaHidaka19}. 
By using the molar 4f specific heat $C_{\mathrm{m}}$ \cite{SupplementalMaterial}, measured on a single crystal from the same batch with a comparable Pr-concentration of $x=0.044$, we evaluate the quadrupolar Grüneisen parameter that follows the expected singular behavior $\Gamma_{\mathrm{u}} \sim H^2/(T^2 \log 1/T)$ for low magnetic fields as shown by the black solid line in Fig.~\ref{fig:fig1}(b). The crossover field of approximately 2\,T is comparable to the crossover field as suggested by the measurement of  $\alpha_{\mathrm{u}}$.

\begin{figure}
       \centering
       \includegraphics[width=0.45\textwidth]{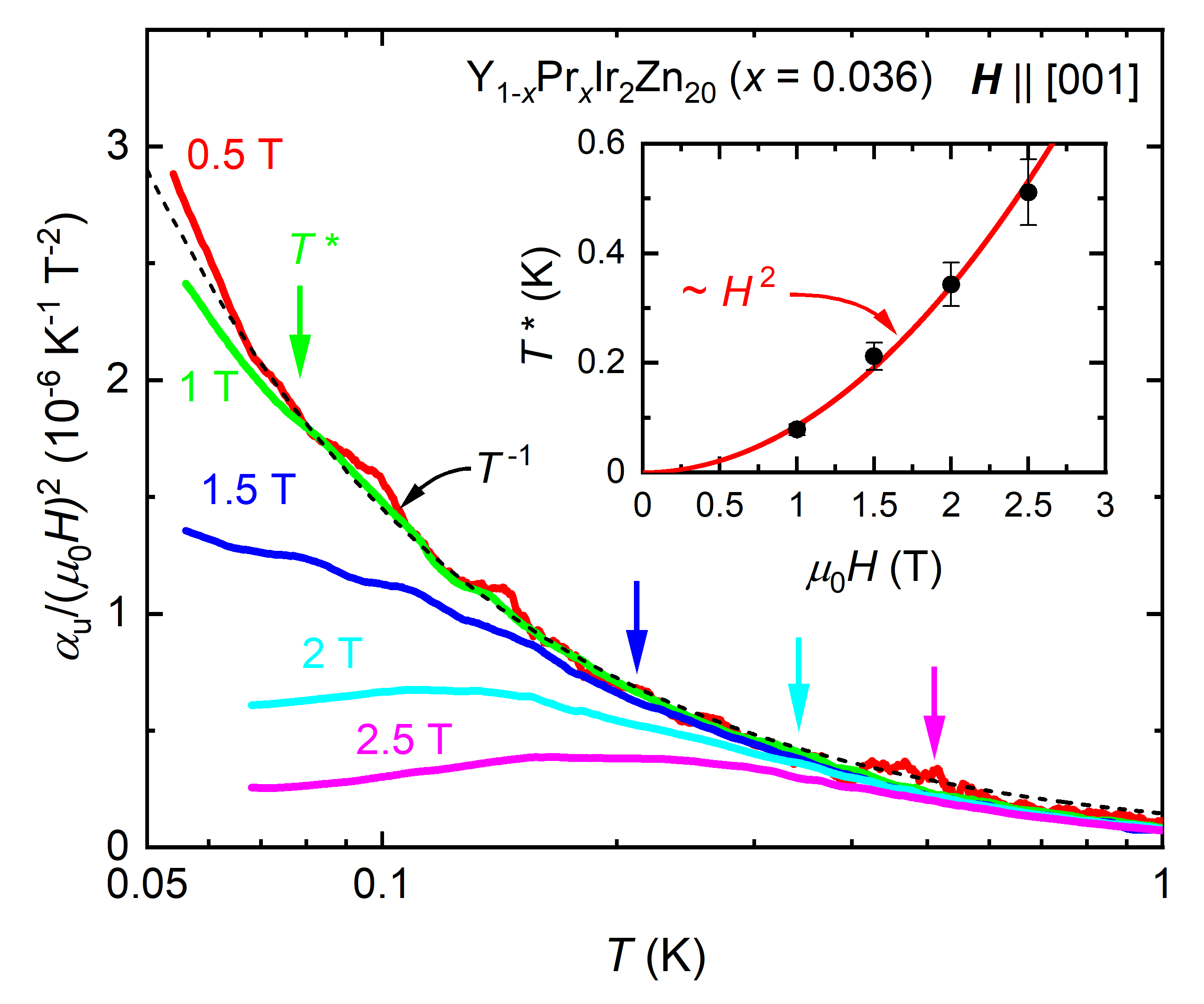}
       \caption{Quadrupolar thermal expansion coefficient normalized to magnetic field $\alpha_{\mathrm{u}}/(\mu_{0}H)^2$ as a function of temperature for magnetic field $\boldsymbol{H}\parallel[001]$. The dashed solid line indicates the $T^{-1}$ single-impurity quadrupole Kondo dependence. Arrows denote a characteristic temperature $T^{*}$, below which deviations from the universal single-impurity quadrupole Kondo behavior arise. The inset displays $T^{*}$ as a function of magnetic field, with the red solid line indicating a $H^{2}$ field dependence.}
       \label{fig:fig4}
 \end{figure}
 
Within the framework of the quadrupolar Kondo model, the deviation from the characteristic quadrupole Kondo behavior for larger magnetic fields can be attributed to both a channel- and a quadrupolar-asymmetry, induced in linear and quadratic order in $H$, respectively.  At small fields, the first effect is expected to dominate and results in a crossover temperature $T^* \sim H^2$, see Ref.~\cite{CoxZawadowski98}, separating non-Fermi and Fermi-liquid behavior. 
In order to specify the origin of the magnetic field induced crossover in highly diluted Y$_{1-x}$Pr$_x$Ir$_2$Zn$_{20}$, Fig.~\ref{fig:fig4} displays the data of $\alpha_{\mathrm{u}}/(\mu_{0}H)^2$ on a logarithmic temperature scale at small magnetic fields up to 2.5\,T. The crossover temperature $T^*$, which is determined as the temperature at which deviations from the universal single-impurity quadrupole Kondo behavior arise, is indicated by an arrow for each magnetic field. The inset displays the characteristic temperature $T^{*}$ as a function of magnetic field, whereby the red solid line indicates a quadratic magnetic field dependence. Indeed, the characteristic temperature $T^*$ estimated from the quadrupolar thermal expansion data can be well scaled with $T^* \sim H^2$. This indicates that the field induced channel-asymmetry is the dominating perturbation at low magnetic field leading to the experimentally found deviations from the quadrupole Kondo behavior at low temperature, which is in very good agreement with the theoretical expectation.

\begin{figure}
\includegraphics[width=0.48\textwidth]{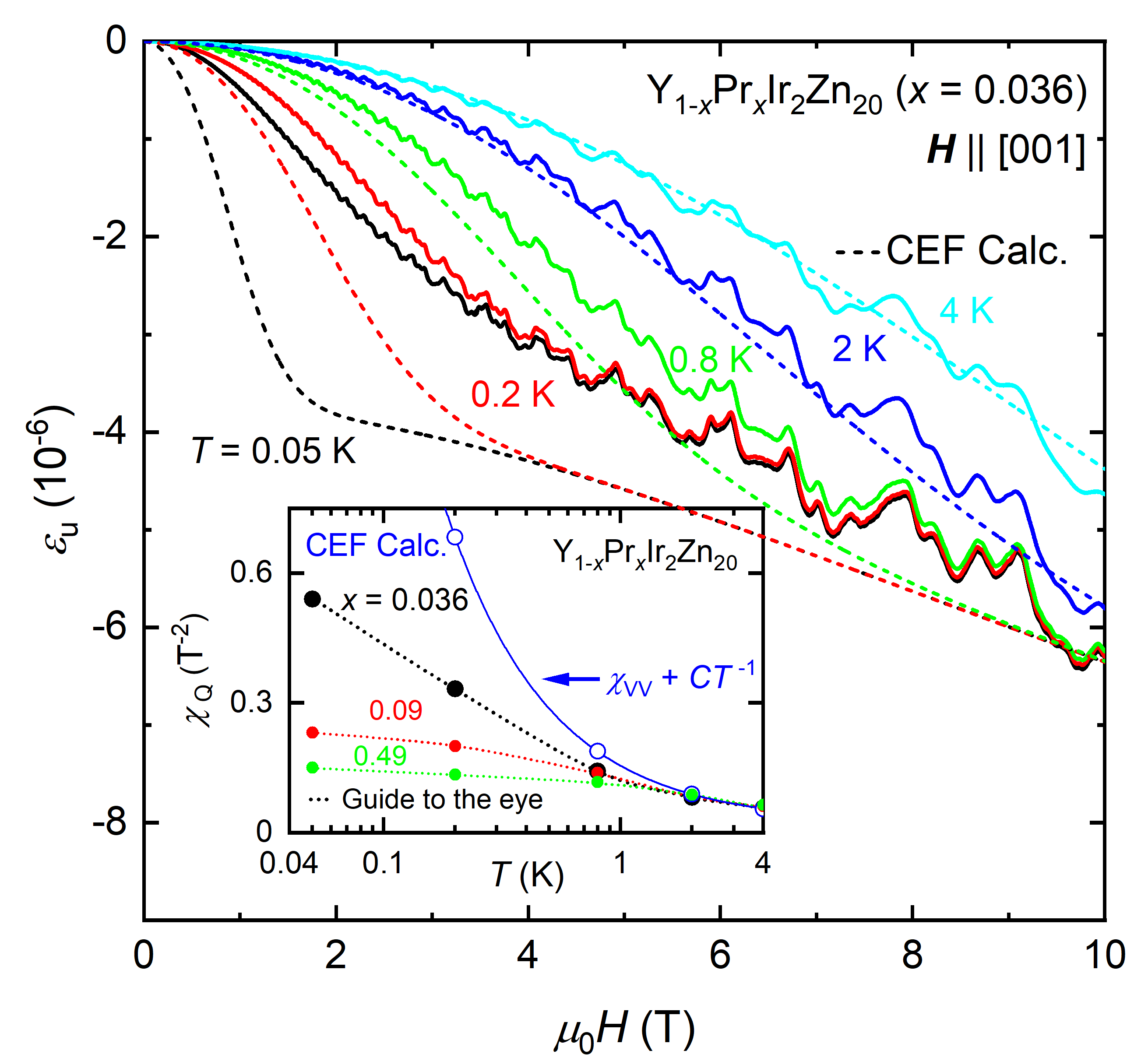}
\caption{Magnetic field variation of the quadrupolar magnetostriction $\varepsilon_{\mathrm{u}}$ for $\boldsymbol H\parallel[001]$ at different temperatures. The dashed lines are CEF calculations. The inset shows the quadrupole susceptibility $\chi_{\mathrm{Q}}$ extracted from the quadratic  dependence $\varepsilon_{\mathrm{u}} \sim n_{\mathrm{Pr}}  \chi_{\mathrm{Q}} H^2$ at small $H$ for three different Pr concentrations $x$. The CEF prediction is shown as a solid blue line, whose temperature dependence is characterized by the superposition of a constant Van Vleck, $\chi_{\mathrm{VV}}$, and a $C/T$ Curie contribution which cannot describe the experimental data at low $T$.
} 
\label{fig:fig2}
\end{figure} 
The results on the quadrupolar thermal expansion coefficient $\alpha_{\mathrm{u}}$ and the quadrupolar Grüneisen parameter $\Gamma_{\mathrm{u}}$, which are indicative of the single-impurity quadrupole Kondo effect, are further corroborated by the quadrupolar magnetostriction $\varepsilon_{\mathrm{u}}$ displayed in Fig.~\ref{fig:fig2}. The quadrupolar magnetostriction coefficient $\varepsilon_{\mathrm{u}}$ is derived from the measurement of the longitudinal $\varepsilon_{\parallel}$ and the volume magnetostriction coefficient $\varepsilon_{\mathrm{B}}$ by using the relation $\varepsilon_{\mathrm{u}}=\sqrt{3}(\varepsilon_{\parallel}-\varepsilon_{\mathrm{B}}/3)$ for $\boldsymbol{H}\parallel{[001]}$. The data of $\varepsilon_{\parallel}$ and $\varepsilon_{\mathrm{B}}$ used for the calculation of $\varepsilon_{\mathrm{u}}$ is provided in the SM~\cite{SupplementalMaterial}. By analyzing the initial quadratic field dependence of $\varepsilon_{\mathrm{u}} \sim n_{\rm Pr} \chi_{\mathrm{Q}} H^2$~\cite{SupplementalMaterial}, we extract the quadrupolar susceptibility $\chi_\mathrm{Q}$ that is shown for various Pr doping $x$ in the inset of Fig.~\ref{fig:fig2}. The CEF calculation, that predicts a Curie-like $1/T$ temperature dependence on top of a constant Van Vleck contribution, can only capture the behavior of $\chi_\mathrm{Q}$ at elevated temperature. At low $T$ and for the lowest doping concentration $x = 0.036$ our results are consistent with a logarithmic temperature dependence of $\chi_{\mathrm{Q}}$. For higher doping, the susceptibility is suppressed at low temperatures indicating a quenching of the quadrupolar Kondo effect, likely induced by interactions between Pr$^{3+}$ ions.

\begin{figure}
\includegraphics[width=0.5\textwidth]{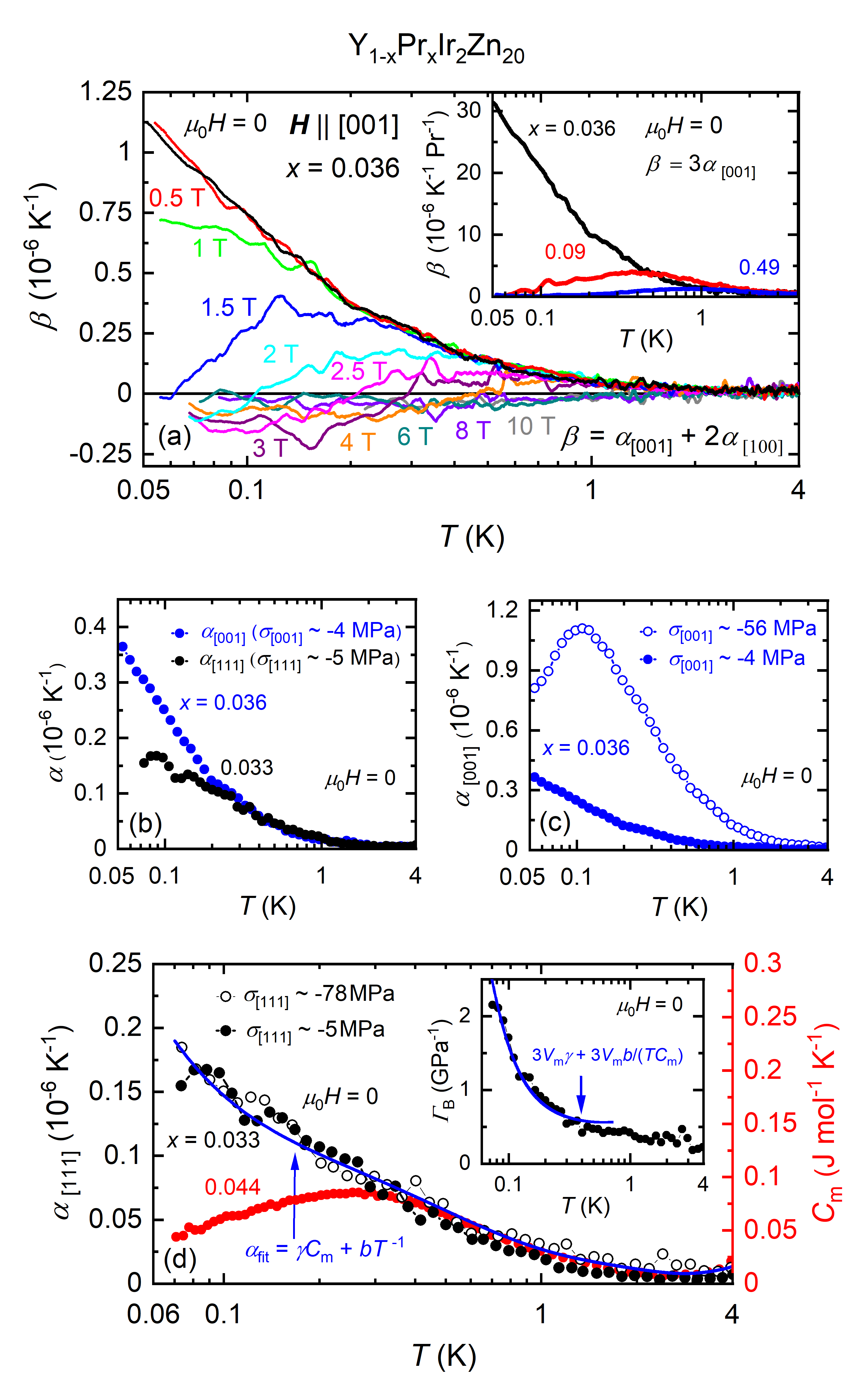}
\caption{ (a) Volume thermal expansion coefficient $\beta$ as a function of temperature at various magnetic fields $\boldsymbol H\parallel[001]$. The inset shows $\beta$ at zero magnetic field for three different Pr-concentrations $x$. (b) Temperature dependencies of the uniaxial thermal expansion coefficients  $\alpha_{[001]}$ and $\alpha_{[111]}$ measured with the miniaturized dilatometer at zero magnetic field. (c) Temperature dependencies of $\alpha_{[001]}$ measured with the miniaturized (closed circles) and the uniaxial stress dilatometer (open circles).  (d) Temperature dependencies of $\alpha_{[111]}$ obtained by the miniaturized (filled circles) and the uniaxial stress dilatometer (open circles) together with specific heat data, partially taken from Ref.~\cite{YamaneOnimaru18}.The blue solid line is a fit to $\alpha_{[111]}$ using $\alpha_{\mathrm{fit}}=\gamma C_{\mathrm{\mathrm{m}}}+bT^{-1}$. The inset displays the temperature dependence of the bulk Grüneisen parameter $\Gamma_{\mathrm{B}}$.
}
\label{fig:fig3}
\end{figure}

We now turn to the discussion of the volume thermal expansion coefficient $\beta = \partial_T \varepsilon_{\mathrm{B}}$. In Fig.~\ref{fig:fig3}(a), $\beta$ is shown for the highly diluted single crystal with $x=0.036$. Remarkably, the thermal expansion coefficient at zero field increases down to lowest temperatures. This continuous increase is quenched by the application of a field and, for $\mu_0 H \gtrsim 6$ T, the volume thermal expansion practically vanishes. Similarly, $\beta$ is also suppressed by larger doping as shown in the inset for zero field. 
The continuous increase of $\beta$ on cooling at zero field and low doping is unexpected at first sight as the influence of the quadrupolar moments should in lowest order average out in the bulk thermal expansion. A parasitic signal could potentially arise from the presence of a small uniaxial stress from the sample holder that is on the order of a few MPa in case of the miniaturized dilatometer and breaks the isotropy of the environment. In order to clarify this influence, we compare in Fig.~\ref{fig:fig3}(b) the uniaxial thermal expansion along crystallographic $[001]$ and $[111]$ directions measured on two single crystals with almost the same doping level. While a uniaxial stress along $[111]$ only activates the bulk strain $\varepsilon_{\mathrm{B}}$, a uniaxial stress along $[001]$ induces both bulk and quadrupolar strains where the latter couples linearly to the $\Gamma_3$ ground state doublet of the Pr$^{3+}$ ions. At very low temperature, we find indeed a difference between $\alpha_{[001]}$ and $\alpha_{[111]}$ that we attribute to the presence of a small uniaxial stress exerted by the dilatometer.

To demonstrate the tremendous effect of uniaxial stress along [001], we performed a complementary measurement using a uniaxial stress dilatometer with an applied uniaxial stress approximately fifteen times larger than in the miniaturized dilatometer, see Fig.~\ref{fig:fig3}(c). Under higher uniaxial stress, the divergence of $\alpha_{[001]}$ is suppressed and a maximum forms at around 0.11\,K, suggesting the breakdown of the two-channel Kondo effect due to the linear in strain splitting of the $\Gamma_3$ ground state doublet.
 
Figure \ref{fig:fig3}(d) displays the temperature dependence of $\alpha_{[111]}$ measured with both the miniaturized and the uniaxial stress dilatometer at zero magnetic field. As there is no notable difference between the two data sets, we conclude that the divergent temperature dependence of $\alpha_{[111]}$ is in fact an intrinsic volume effect and not artificially induced by the  uniaxial stress applied externally along the measurement direction. This result is striking as it is markedly different from the molar specific heat of the 4f electrons $C_{\mathrm{m}}$ as shown by the red symbols. 
The thermal expansion is well described by a fit $\alpha_{\mathrm{fit}} = \gamma C_{\mathrm{m}} + b/T$ (blue line) that assumes two contributions. The first obeys Gr\"uneisen scaling with a constant $\gamma$ and the second contribution diverges like $1/T$, which is reminiscent of the behavior of $\alpha_{\mathrm{u}}/H^2$ shown in the inset of Fig.~\ref{fig:fig1}(b). Consequently, the bulk Gr\"uneisen ratio $\Gamma_{\mathrm{B}} = V_{\mathrm{m}}3\alpha_{[111]}/C_{\mathrm{m}} \approx 3V_{\mathrm{m}}\gamma + 3V_{\mathrm{m}}b/(T C_{\mathrm{m}})$ exhibits a  divergence as a function of temperature similar to $\Gamma_{\mathrm{u}}/H^2$ shown in Fig.~\ref{fig:fig1}(b).  

In the following, we speculate on the origin of this surprising finding of an intrinsic bulk thermal expansion that increases down to lowest temperatures. 
The elastic coupling $g_{\Gamma_3}$ to the non-Kramers $\Gamma_3$ doublet leads in second order perturbation theory 
to the following correction to the elastic Hamiltonian
\begin{align} \label{HCorr}
\delta \mathcal{H} = - \frac{g_{\Gamma_3}^2 \chi_{\mathrm{Q}}}{2} \sum_\alpha 
(\varepsilon^2_{\mathrm{u}}(\vec r_\alpha) + \varepsilon^2_{\mathrm{v}}(\vec r_\alpha)).
\end{align}
Here, $\varepsilon_{\mathrm{u/v}}(\vec r_\alpha)$ are the two components of the local quadrupolar strain doublet at the position $\vec r_\alpha$ of the Pr$^{3+}$ ion with index $\alpha$. After averaging over Pr disorder this Hamiltonian recovers the renormalization of the quadrupolar elastic constant $\delta c_{\mathrm{u}} = -n_{\mathrm{Pr}} g_{\Gamma_3}^2 \chi_{\mathrm{Q}}$ whose temperature dependence was confirmed by elastic constant measurements \cite{YanagisawaHidaka19}. 
Treating the Hamiltonian of Eq.~\eqref{HCorr} in perturbation theory, one obtains a correction to the free energy density of the form $\delta f \sim \frac{\delta c_u}{2} \langle \varepsilon_{\mathrm{u/v}}^2 \rangle$. Here, a finite expectation value $\langle \varepsilon_{\mathrm{u/v}}^2 \rangle$ might arise either from dynamic zero-point fluctuations of the quadrupolar strain or from static strain disorder \cite{SupplementalMaterial}. This results in a contribution to the bulk thermal expansion $\delta \beta = \partial_p \partial_T \delta f = 
- \frac{1}{2} (\partial_p n_{\mathrm{Pr}} g_{\Gamma_3}^2 \langle \varepsilon_{\mathrm{u/v}}^2 \rangle) \partial_T \chi_{\mathrm{Q}}$. The  divergence $\delta \beta \sim 1/T$ originating from $\chi_Q \sim \log 1/T$ is consistent with the observations in Fig.~\ref{fig:fig3}(d).

Hence, dynamic as well as static strain fluctuations can in principle explain the observed singular behavior in $\beta$. However, both scenarios also imply a contribution to the specific heat $\delta C_{\mathrm{m}} \sim 1/T$, which so far has not been successfully identified. A possible reason is a small prefactor of $\delta C_{\mathrm{m}}$, or, equivalently, a large effective Gr\"uneisen parameter $\delta \Gamma_{\mathrm{B}} = \delta \beta/\delta C_{\mathrm{m}}$ characterizing the hydrostatic pressure dependence of $n_{\mathrm{Pr}} g_{\Gamma_3}^2 \langle \varepsilon_{\mathrm{u/v}}^2 \rangle$. This dependence can arise from elastic anharmonicities, e.g., via the elastic coupling $g_{\Gamma_3}(p)$ or, in case of strain disorder, from a response of $n_{\mathrm{Pr}}  \langle \varepsilon_{\mathrm{u/v}}^2 \rangle$ to pressure changes. From Fig.~\ref{fig:fig3}(d) we can estimate a lower bound $\Gamma_{\mathrm{B}} > 2.15$ \,GPa$^{-1}$, and with the isothermal compressibility $\kappa_T \approx 0.01$ GPa$^{-1}$, estimated from elastic constant measurements on PrIr$_2$Zn$_{20}$ \cite{IshiiMuneshige11}, this implies a dimensionless Gr\"uneisen parameter $\delta \Gamma_{\mathrm{B}} \kappa_T$ at least of the order of 200 to be compared with a typical value of 1 for metals \cite{AshcroftMermin76}. As elastic anharmonicities are in general unlikely to account for such large values, this suggests static strain disorder to be at its origin.
Hydrostatic pressure will influence the local strain distribution and  might thus affect sensitively the quadrupolar environment of each Pr ions resulting in a large response. 
Local strain fields were also invoked to explain the temperature and field dependence of the specific heat \cite{YamaneOnimaru18} and the elastic constant \cite{YanagisawaHidaka19}, respectively. 
Future nuclear quadrupolar resonance (NQR) measurements might help clarifying this issue.

In this work, we established the symmetrized quadrupolar expansivity, i.e., thermal expansion and magnetostriction, as well as the quadrupolar Gr\"uneisen parameter as a highly sensitive probe to uncover quadrupolar fluctuations. For the example of highly diluted Y$_{1-x}$Pr$_{x}$Ir$_2$Zn$_{20}$ we demonstrated that it gives access to the quadrupolar susceptibility. Our experimental results are in excellent agreement with expectations for the single-impurity quadrupolar Kondo effect and corroborate its emergence in this material. On a more general note, we anticipate the quadrupolar expansivity and quadrupolar Gr\"uneisen parameter to become an important tool also for the investigation of other materials with pronounced quadrupolar correlations, in particular, for materials close to a nematic quantum critical point as discussed for certain cuprate and iron pnictide superconductors~\cite{Auvray19,Worasaran21}.


\begin{acknowledgments}
Long term collaboration with R. K\"uchler as well as helpful discussions with Y. Tokiwa, R. Yamamoto, T. Yanagaisawa and H. Kusunose are gratefully acknowledged. The work in Augsburg was supported by the German Research Foundation (DFG) via the Project No. 107745057 (TRR80). The work in Hiroshima was supported by the Center for Emergent Condensed-Matter Physics (ECMP), Hiroshima University and financially supported by Grants-in-Aid from MEXT/JSPS of Japan [Grant Nos. JP26707017, JP15H05886 (J-Physics), JP15KK0169, JP18H01182, and JP18KK0078]. M.G. is supported by the DFG via the Project No. 422213477 (TRR 288, project A11).
\end{acknowledgments}

\bibliography{YPrIr2Zn20.bib}
\newpage

\onecolumngrid

\newcommand{\MG}[1]{{\color{red} #1}}

\renewcommand{\thefigure}{S\arabic{figure}}
\renewcommand{\theequation}{S\arabic{equation}}
\renewcommand{\thetable}{S\arabic{table}}
\setcounter{equation}{0}
\setcounter{figure}{0}
\setcounter{table}{0}
\begin{center}
\textbf{\large Supplemental Material for \\``Divergent Thermal Expansion and Gr\"uneisen Ratio in a Quadrupolar Kondo Metal''}\\
\vspace{0.5cm}
A. W\"orl, M. Garst, Y. Yamane, S. Bachus, T. Onimaru, and P. Gegenwart
\end{center}

\section{Experimental methods}
This section provides detailed information on the investigated single crystalline samples of Y$_{1-x}$Pr$_x$Ir$_2$Zn$_{20}$. Furthermore, the two different capacitive dilatometer employed in this study are briefly introduced. 
\subsection{Y$_{1-x}$P\lowercase{r}$_{x}$I\lowercase{r}$_2$Z\lowercase{n}$_{20}$ Single Crystals} 
\begin{figure}[b]
       \centering
       \includegraphics[width=0.8\textwidth]{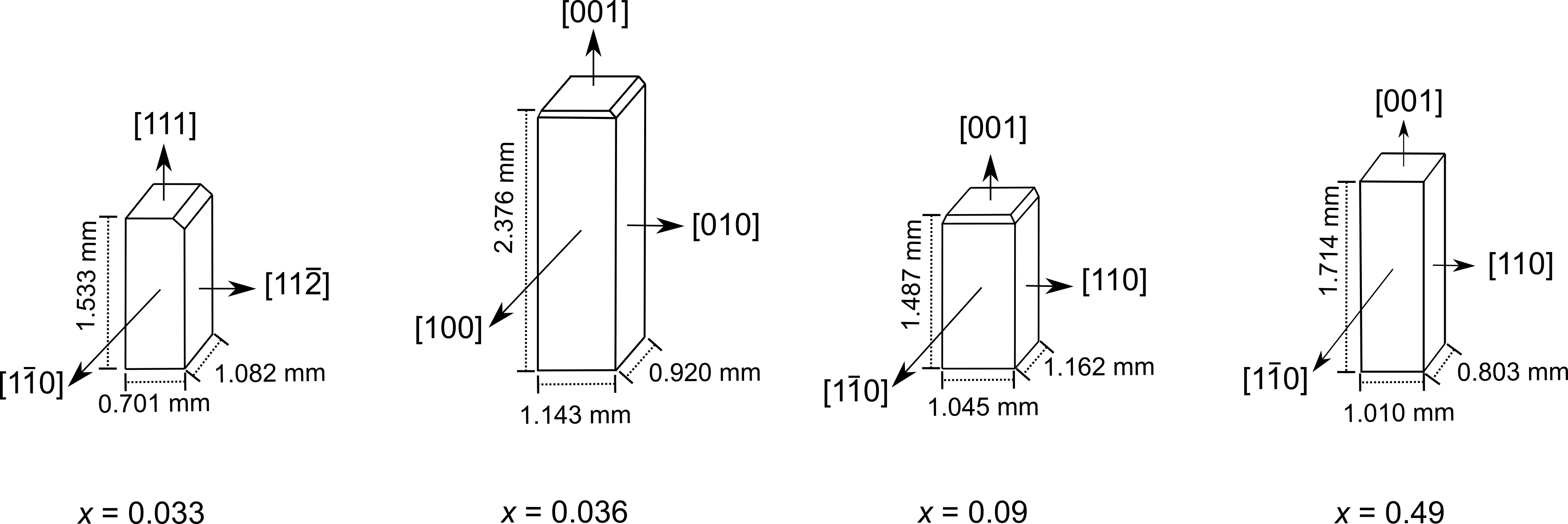}
       \caption{Sketches of the studied Y$_{1-x}$Pr$_x$Ir$_2$Zn$_{20}$ single crystals which illustrate the shape, crystallographic directions and dimensions for each single crystal.}
       \label{fig:figS1}
 \end{figure}

The Y$_{1-x}$Pr$_{x}$Ir$_{2}$Zn$_{20}$ single crystals studied in this work were grown by use of a Zn self-flux technique. For more details on the single crystal growth process, we refer to Ref.~ \cite{YamaneOnimaru18PhysicaB}. In this study, we focus on four differently doped Y$_{1-x}$Pr$_{x}$Ir$_{2}$Zn$_{20}$ single crystals with Pr-concentrations of $x=0.033, 0.036, 0.09$ and $0.49$.  The shape, crystallographic directions and dimensions of each single crystal are illustrated in Fig. \ref{fig:figS1}, whereby the given length values were determined with an accuracy of $\pm 0.001$\,mm.

The Pr concentrations $x$ of the single crystalline sample with $x=0.49$ was determined by use of electron-probe microanalysis (EPMA), whereby the uncertainty in the concentration is $\Delta x \pm 0.06$. As EPMA turned out as not sensitive enough for characterizing the higher diluted samples with $x=0.033$, $x=0.036$ and $x=0.09$, their magnetization was measured at 1.8\,K and 1\,T and the respective Pr-concentration estimated by comparing the experimental value with the theoretical one suggested by the crystal electric field (CEF) calculation. These are the same characterization procedures as already applied in previous studies \cite{YamaneOnimaru18, YamaneOnimaru18PhysicaB, YanagisawaHidaka19}. The single crystal with $x=0.044$ on which the specific heat presented in Fig. 4(d) of the main paper was measured, is from the identical sample batch as the two highly diluted single crystals with $x=0.033$ and $x=0.036$, on which the thermal expansion was obtained. The uncertainty in the Pr concentration of the three highly diluted single crystals with $x=0.033$, $x=0.036$ and $x=0.044$ is estimated at $\Delta x = \pm 0.01$. The Pr content of these three single crystalline samples can therefore be considered as the same within the estimated error of the experimental technique used for the determination of the Pr concentration. The uncertainty in the Pr concentration of the single crystalline sample with $x=0.09$ is estimated at $\Delta x = \pm 0.02$.

The previously published values for the residual resistance ratio ($RRR$)
of Y$_{1-x}$Pr$_{x}$Ir$_{2}$Zn$_{20}$ imply that the highly diluted single crystals are of very high quality. The  $RRR$ of the highly diluted sample with $x=0.036$ is estimated at 160, which is the value determined previously for a sample ($x=0.044$) from the same batch \cite{YamaneOnimaru18PhysicaB}. By increasing the Pr-concentration the $RRR$ decreases continuously, which can be explained by increasing disorder. The $RRR$ of the $x=0.49$ sample is estimated at only 8, which is the previously reported value of a single crystal $(x=0.44)$ from the same batch \cite{YamaneOnimaru18PhysicaB}.  

\subsection{Dilatometry} 

Ultrahigh resolution relative length change measurements were carried out by employing a miniaturized capacitive dilatometer \cite{KuechlerWoerl17}. The dilatometer is made of a copper beryllium alloy and has super compact dimensions of $15\,\mathrm{mm}\times 14\,\mathrm{mm} \times 14.7\,\mathrm{mm}$~\cite{KuechlerWoerl17}.
This allows for its rotation inside the small sample space of a dilution refrigerator and enables relative length change measurements both parallel and perpendicular to magnetic field.  

The relative length change $\Delta L$ of the sample was determined by measuring the capacitance of the dilatometer as a function of temperature or magnetic field.  To convert the respective capacitance values into relative length changes, we used the formula by Pott and Schefzyk \cite{PottSchefzyk83}, which takes account of the maximal adjustable capacitance value $C_{\mathrm{max}}$ of the dilatometer. This corrects a small error which originates from non perfectly parallel capacitor plates. For the miniaturized capacitive dilatometer, the maximal adjustable capacitance value is $C_\mathrm{max}=50$\,pF. The capacitor plate radius of the miniaturized dilatometer is 5\,mm \cite{KuechlerWoerl17}.

As the sample is clamped by four leaf springs into the dilatometer, a small inevitable force acts on the sample in measurement direction. This force varies with the adjusted capacitance value and equals $4$\,N when the capacitive dilatometer is operated at a capacitance value of $20\,$pF \cite{KuechlerWoerl17}. As measurements on the different samples were performed at capacitance values ranging between $19.8$\,pF and $20.7$\,pF, we simply assume that the same force of $4$\,N acts on all investigated samples. The uniaxial stress on the sample is estimated as $\sigma = F/A$, where $F=4$\,N is the force exerted on the sample by the miniaturized capacitive dilatometer and $A$ and the surface area of the sample. As the sample shapes are close to rectangular (cf. Fig. \ref{fig:figS1}), we approximately calculate the surface area by the lengths of the two sides perpendicular to the measurement direction.

To quantify the impact of uniaxial stress on the measurement results, we carried out additional thermal expansion measurements on the $x=0.036$ (along [001]) and on the $x=0.033$ (along [111]) single crystals by use of a uniaxial stress dilatometer~\cite{KuechlerStingl16}. This dilatometer is also manufactured from a copper beryllium alloy and its working principle is identical to the previously described miniaturized capacitive dilatometer, with the important difference that its leaf springs are much stronger, therefore exerting a higher force on the sample. Furthermore, it has a larger capacitor plate radius of 7\,mm \cite{KuechlerStingl16}. Relative length changes were calculated by the same procedure as discussed above, whereby the maximal adjustable capacitance value of the uniaxial stress dilatometer is $C_{\mathrm{max}}=90$\,pF. The force acting on the sample in measurement was estimated by means of the previously published relation between capacitance and force \cite{KuechlerStingl16}, which yields approximately 59.1\,N and 58.7\,N for the measurements on the $x=0.033$ and $x=0.036$ samples, respectively. After the measurement under high uniaxial stress the appearance of the two investigated samples was checked. The $x=0.036$ sample did not show any indication of damage. The $x=0.033$ sample exhibited small cracks. 

The thermal expansion and magnetostriction curves presented in the main text are background corrected, whereby the cell effect of the dilatometer was determined by measuring relative length changes of the capacitive dilatometer with a Cu specimen placed inside.

\section{Theoretical considerations}

\subsection{Crystal Electric Field Hamiltonian}
 
The fully localized quadrupole moments of each Pr$^{3+}$ ion ($J=4$) is described by the crystal electric field (CEF) Hamiltonian \cite{LeaLeask62} for the $T_{d}$ point group of the cubic crystal symmetry
\begin{equation}
\mathcal{H}_{\mathrm{CEF}}=W\left[\frac{x}{60}(O_4^0 +5O_4^4)+\frac{1-|x|}{1260}(O_6^0-21O_6^4)\right]-g_{J}\mu_{\mathrm{B}}\boldsymbol{J}\boldsymbol{H}
\label{ref:EqS1}
\end{equation}
where $O_k^q$ are the Stevens operators given in terms of the angular momentum operator $\boldsymbol{J}$. The CEF parameters
used to describe the thermal expansion and magnetostriction of Y$_{1-x}$Pr$_x$Ir$_2$Zn$_{20}$ $(x=0.036)$ are $W=-1.50$\,K and $x=0.537$. The same parameters were used previously in order to account for the temperature and field dependence of entropy for a sample with a comparable Pr concentration of $x=0.044$~\cite{YamaneOnimaru18AIP}. We note that the value $W=-1.50$\,K differs slightly from the value $W=-1.22$\,K~\cite{IwasaHiroki13}, which was suggested by inelastic neutron scattering on pure PrIr$_2$Zn$_{20}$.  Possible explanation for the difference between the Y diluted system and pure system is a strengthening of the CEF effect due the smaller ionic radius of Y as compared to Pr. The second term of the Hamiltonian is the Zeeman term, where $g_{J}$ is the Land\'e factor and $\mu_{\mathrm{B}}$ the Bohr magneton. 

In the absence of a magnetic field $\boldsymbol{H}$, the groundstate of Eq.~\eqref{ref:EqS1} is given by a degenerate non-Kramers doublet with quadrupolar $\Gamma_3$ symmetry,
\begin{align}
|\Gamma_3,+ \rangle &= 
\sqrt{\frac{7}{24}} \Big(|4 \rangle - \sqrt{\frac{10}{7}} |0 \rangle + |-4\rangle \Big),\\
|\Gamma_3,- \rangle &= 
\sqrt{\frac{1}{2}} \Big(|2 \rangle  + |-2\rangle \Big),
\end{align}
with the spin eigenstates $| m \rangle$ and $m \in \{ -4,-3,...,3,4\}$. These states can be associated with the Hilbert-space of a pseudospin operator $\boldsymbol{\sigma}_{\Gamma_3}$. An important role is played by the Stevens operators $O^0_{2} \propto 2 J_z^2 - J_x^2-J_y^2$ and $O^2_{2} \propto J_x^2 - J_y^2$. After projection onto the degenerate groundstate doublet, these operators can be identified with components of the pseudospin operator $\boldsymbol{\sigma}_{\Gamma_3}$. Consequently, their thermal expectation value with respect to the non-Kramers doublet vanishes at zero magnetic field, $\langle O^q_{2} \rangle = 0$ for $q=0,2$. We will be particularly interested in a finite magnetic field applied along the cubic crystal axis $[ 001]$ that splits the doublet and leads to a finite expectation value of $\langle O^0_{2} \rangle \sim \mathcal{O}(H^2)$.

\subsection{Quadrupolar Kondo coupling to the conduction electrons}

The local pseudospin $\boldsymbol{\sigma}_{\Gamma_3}$ will couple with a Kondo exchange coupling $J_K$ to the conduction electrons via two channels corresponding to the two electronic spin-configurations potentially realizing the two-channel Kondo (2CK) effect as suggested by Cox \cite{Cox87}. In the dilute limit of a few Pr$^{3+}$ ions the interaction of each pseudospin with the itinerant electrons might then be governed at low temperature by the non-Fermi liquid fixed point of the two-channel Kondo model characterized by a residual entropy of $\frac{k_B}{2} \log 2$ per ion. Close to this fixed-point, the quadrupolar pseudospin susceptibility $\chi_Q$, that characterizes the response of the pseudospin to an infinitesimally small pseudomagnetic field $\boldsymbol{H}_{\Gamma_3}$, 
is  logarithmically divergent with temperature, $\chi_Q \sim \log \frac{1}{T}$.
A pseudomagnetic field is generated when the quadrupolar symmetry of the Pr$^{3+}$ environment gets broken. This is, for example, achieved by either a suitably applied elastic strain or by the application of a magnetic field, where $H_{\Gamma_3} \sim H^2$ for small fields $H$. 

\subsection{Elastic coupling to the local quadrupole moment}

The Stevens operators can locally couple to the strain tensor of the crystal lattice $\varepsilon_{ij}(\vec r)$,
\begin{align}
\label{Hstrain}
\mathcal{H}_{\rm strain} = \int d\vec r \frac{1}{2} \varepsilon_{ij}(\vec r) C_{ijkl} \varepsilon_{kl}(\vec r) 
- g_{\Gamma_3} \sum_\alpha  \left[
O^0_{2, \alpha} \varepsilon_u(\vec r_\alpha)
+O_{2, \alpha}^2 \varepsilon_{\mathrm{v}}(\vec r_\alpha) \right] - g_{\Gamma_1} \sum_\alpha O^0_{0, \alpha} \varepsilon_B(\vec r_\alpha)
\end{align}
where $C_{ijkl}$ is the elastic constant matrix and $O^q_{k, \alpha}$ are the Stevens operator associated with the Pr$^{3+}$ ion enumerated by the index $\alpha$ and located at position $\vec r_\alpha$. We have assumed for simplicity that the elastic coupling constants, $g_{\Gamma_3}$ and $g_{\Gamma_1}$, are the same for all $\alpha$. The Stevens operators $O^0_{2, \alpha}$ and $O^2_{2, \alpha}$ generate a local quadrupolar stress, whereas $O^0_{0, \alpha} = 1$ generates a local isotropic stress. The crystal responds to the local stresses leading to a homogeneous and inhomogeneous strain distribution.

\subsubsection{Macroscopic strains induced by the Pr ions}

In the following, we focus on the homogeneous macroscopic strain denoted by $\varepsilon_{ij} = \frac{1}{V} \int d\vec r \varepsilon_{ij}(\vec r)$.  The macroscopic eigenstrains and the corresponding elastic constants in a cubic system are listed in Table \ref{ref:tab1} where we adopt the convention of the previous literature on Y$_{1-x}$Pr$_x$Ir$_2$Zn$_{20}$.

\begin{table}[htb]
\centering
\begin{tabular}{ccc}\toprule
Symmetry & Symmetry Strains  & Symmetry Elastic Constants 
\\ \midrule
$\Gamma_1$ & $\varepsilon_{\mathrm{B}}=\varepsilon_{xx}+\varepsilon_{yy}+\varepsilon_{zz}$ & $(c_{11}+2c_{12})/3$ \\[0.02cm]
$\Gamma_3$ & $\varepsilon_{\mathrm{u}}=(2\varepsilon_{zz}-\varepsilon_{xx}-\varepsilon_{yy})/\sqrt{3}$ & $(c_{11}-c_{12})/2$\\
& $\varepsilon_{\mathrm{v}}=\varepsilon_{xx}-\varepsilon_{yy}$ & $(c_{11}-c_{12})/2$ \\[0.02cm]
$\Gamma_5$ & $\varepsilon_{xy}, \varepsilon_{xz}, \varepsilon_{yz}$ & $4c_{44}$ \\[0.02cm]
\bottomrule
\end{tabular}
\caption{Symmetries, respective symmetry strains and symmetry elastic constants for cubic symmetry.}
\label{ref:tab1}
\end{table}

Minimizing Eq.~\eqref{Hstrain} with respect to the macroscopic strains we obtain
\begin{align} 
\varepsilon_{\mathrm{B}} &= \frac{g_{\Gamma_1} n_{\rm Pr}}{(c^0_{11}+2 c^0_{12})/3},
\qquad
\varepsilon_{\mathrm{u}} = \frac{g_{\Gamma_3}}{(c^0_{11}-c^0_{12})/2} \frac{1}{V}\sum_\alpha \left\langle O^0_{2,\alpha} \right\rangle,
\qquad
\varepsilon_{\mathrm{v}} = 
\frac{g_{\Gamma_3}}{(c^0_{11}-c^0_{12})/2} \frac{1}{V}\sum_\alpha \left\langle O^2_{2,\alpha} \right\rangle,
\label{ref:Eq3}
\end{align}
where $n_{\mathrm{Pr}}$ is the Pr density.
The macroscopic strain induced by the Pr$^{3+}$ ions is related to the thermal expectation values $\langle O^q_{k,\alpha} \rangle$, and  $(c^0_{11}+2c^0_{12})/3$ and $(c^0_{11}-c^0_{12})/2$ are the background elastic constants with $\Gamma_1$ and $\Gamma_3$ symmetry, respectively. Whereas the expectation value $\left\langle O^0_{0,\alpha} \right\rangle = 1$ is trivial, the ones in the $\Gamma_3$ channel are governed by the thermal occupation of the non-Kramers doublet. Importantly, these expectation values $\langle O^{q}_{2,\alpha}\rangle$ will in general depend themselves on both elastic coupling $g_{\Gamma_3}$ and the Kondo coupling $J_K$. Moreover, they will also depend on the local strain distribution at the site $\vec r_\alpha$. Only after averaging over this strain disorder one obtains
\begin{align}
\varepsilon_{\mathrm{B}} &\approx \frac{n_{\mathrm{Pr}} g_{\Gamma_1}}{(c^0_{11}+2 c^0_{12})/3}  \langle O^0_{0} \rangle,
\qquad
\varepsilon_{\mathrm{u}} \approx \frac{n_{\mathrm{Pr}}{g_{\Gamma_3}}}{(c^0_{11}-c^0_{12})/2} \langle O^0_2  \rangle,
\qquad
\varepsilon_{\mathrm{v}} \approx \frac{n_{\mathrm{Pr}}{g_{\Gamma_3}}}{(c^0_{11}-c^0_{12})/2} \langle O^2_2 \rangle,
\label{ref:Eq4}
\end{align}
with the Pr density $n_{\mathrm{Pr}}$. 

In general, a macroscopic strain along a certain direction specified by the unit vector $\hat n$ is given by $\hat n_i \varepsilon_{ij} \hat n_j$.
In particular, the strains along the cubic directions [100], [010] and [001] as well as along the diagonal [111] are related to the eigenstrains via 
\begin{align}
\frac{\Delta L}{L}\bigg\rvert_{[001]} &=\varepsilon_{zz}=\frac{1}{3}\varepsilon_{\mathrm{B}}+\frac{1}{\sqrt{3}}\varepsilon_{\mathrm{u}},
\label{ref:Eq5}
\\
\frac{\Delta L}{L}\bigg\rvert_{[100]} &=\varepsilon_{xx}=\frac{1}{3}\varepsilon_{\mathrm{B}}-\frac{1}{2\sqrt{3}}\varepsilon_{\mathrm{u}}+\frac{1}{2}\varepsilon_{\mathrm{v}},
\label{ref:eq6}
\\
\frac{\Delta L}{L}\bigg\rvert_{[010]} &=\varepsilon_{yy}=\frac{1}{3}\varepsilon_{\mathrm{B}}-\frac{1}{2\sqrt{3}}\varepsilon_{\mathrm{u}}-\frac{1}{2}\varepsilon_{\mathrm{v}},
\label{ref:Eq7}
\\
\frac{\Delta L}{L}\bigg\rvert_{[111]} &=\frac{1}{3}\varepsilon_{\mathrm{B}} + \frac{2}{3} \Big(\varepsilon_{xy} + \varepsilon_{zx} + \varepsilon_{yz} \Big).
\label{ref:Eq111}
\end{align}
We singled out the [001] direction along which the magnetic field will be applied. In the absence of any internal or external strains that break additional symmetries, we expect that the shear strains vanish $\varepsilon_{xy} = \varepsilon_{zx} = \varepsilon_{yz} = 0$. In this case the strain along [111] is only sensitive to the bulk strain $\varepsilon_B$. In case the cubic symmetry is maintained for zero magnetic field, the $\Gamma_3$ doublet remains degenerate and $\langle O^q_2 \rangle =0$. 
In turn, this implies vanishing quadrupolar strains $\varepsilon_{\mathrm{u}} = \varepsilon_{\mathrm{v}} = 0$, so that only the isotropic bulk strain $\varepsilon_{\mathrm{B}}$ will contribute to a finite expansion. A magnetic field along $\boldsymbol{H}\parallel [001]$ leads to a tetragonal distortion via the elastic coupling resulting in a finite value for $\varepsilon_{\mathrm{u}}$. The strain $\varepsilon_{\mathrm{v}}$ still remains zero implying equivalent strains $\varepsilon_{xx}$ and $\varepsilon_{yy}$ within the plane perpendicular to the applied magnetic field.

Evaluating the thermal average in Eqs.~\eqref{ref:Eq4} with the CEF Hamiltonian Eq.~\eqref{ref:EqS1} in zeroth order in the Kondo coupling $J_K$ and treating the elastic coupling $g_{\Gamma_3}$ in Eq.~\eqref{Hstrain} in a mean-field approximation, one obtains a characteristic dependence of the uniaxial strains on temperature and magnetic field that are shown as dashed lines in Fig. 1 of the main text. Here, we used $(c_{11}^{0}-c_{12}^{0})/2=52.771$\,GPa \cite{YanagisawaHidaka19} and $N_{\mathrm{Pr}}=0.101 \times 10^{27}$\,m$^{-3}$.

\begin{figure}[t]
\includegraphics[width=0.48\textwidth]{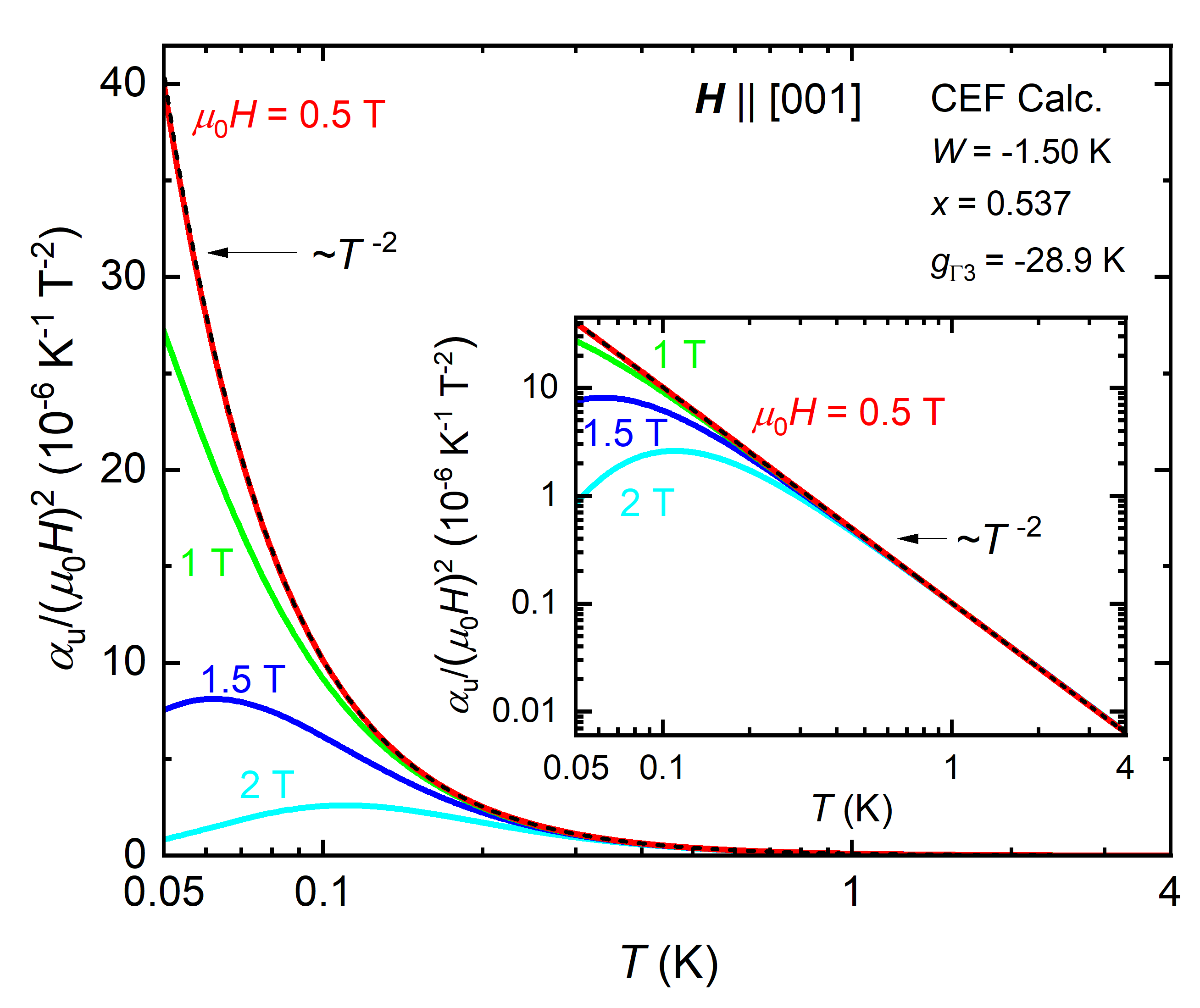}
\caption{Quadrupolar thermal expansion $\alpha_u = \partial_T \varepsilon_u$ divided by $(\mu_0 H)^2$ versus temperature as predicted by the crystal field Hamiltonian of Eq.~\eqref{ref:EqS1}. The simulation parameters are the ones of single crystal with Y$_{1-x}$Pr$_x$Ir$_2$Zn$_{20}$ $(x=0.036$).}  
\label{fig:figSCEF}
\end{figure}

In Fig.~\ref{fig:figSCEF} the CEF prediction for the quadrupolar thermal expansion, $\alpha_u = \partial_T \varepsilon_u$, is shown indicating that at small fields its temperature dependence is governed by the Curie-Weiss behavior, $\alpha_u/H^2 \sim \partial_T \chi_Q \sim 1/T^2$.

\subsubsection{Influence of local strains}

In the presence of finite local strains induced by crystal defects or inhomogeneities, the symmetry of the local Pr environment will be broken which splits the non-Kramers doublet. This results in a finite 
expectation value of the quadrupole moments, that for small strains is given by $\langle O^0_{2,\alpha} \rangle \approx \chi_Q g_{\Gamma_3} Q^u_{ij} \varepsilon^{\rm loc}_{ij}(\vec r_\alpha)$ and $\langle O^2_{2,\alpha} \rangle \approx \chi_Q g_{\Gamma_3} Q^v_{ij} \varepsilon^{\rm loc}_{ij}(\vec r_\alpha)$. Using Eq.~\eqref{ref:Eq3} we then obtain a contribution to the macroscopic quadrupolar strains
\begin{align} \label{QuadrupolarStaticStrain2}
\varepsilon_{u/v} = \frac{g_{\Gamma_3} \chi_Q g_{\Gamma_3} Q^{u/v}_{ij} }{(c^0_{11}-c^0_{12})/2} \frac{1}{V}\sum_\alpha  \varepsilon^{\rm loc}_{ij}(\vec r_\alpha)
\end{align}
where we abbreviated $Q^v_{ij} =  \delta_{ix}\delta_{jx} - \delta_{iy}\delta_{jy}$, and $Q^u_{ij} =  \frac{1}{\sqrt{3}} (2 \delta_{iz}\delta_{jz} - \delta_{ix}\delta_{jx}- \delta_{iy}\delta_{jy})$. After disorder averaging, this contribution vanishes by definition because $\int d\vec r \varepsilon^{\rm loc}_{ij}(\vec r) = 0$. However, for a given strain distribution this might yield a finite contribution that scales linearly with the Pr density $\varepsilon_{u/v} \sim n_{\rm Pr} \chi_Q$ and breaks the cubic crystal symmetry even at zero magnetic field.

The Pr ions themselves will also give rise to local strain fields. In general, the local strains, $\varepsilon^{\rm loc}_{ij} = \frac{1}{2} (\partial_i u_j + \partial_j u_i)$, are represented in terms of the displacement vector $u_i$. Minimizing Eq.~\eqref{Hstrain} with respect to $u_i$ we obtain for the displacement vector in Fourier space $u_i(\vec k)$
\begin{align}
- C_{ijkl} k_j k_k u_l(\vec k) =  \sum_\alpha e^{-i \vec k \vec r_\alpha}  
i k_j \Big[g_{\Gamma_3} Q^v_{ij} \langle O^0_{2,\alpha} \rangle + g_{\Gamma_3} Q^u_{ij} \langle O^2_{2,\alpha} \rangle + g_{\Gamma_1} \delta_{ij}  \Big] 
\end{align}
where we used that $\langle O^0_{0,\alpha} \rangle = 1$. In case the last term dominates and the quadrupolar expectation values $\langle O^q_{2,\alpha} \rangle$ can be neglected, the Pr$^{3+}$ ions induce a local strain field given by
\begin{align}
u_l(\vec k) = - i g_{\Gamma_1} D^{-1}_{ln}(\vec k) k_n \sum_\alpha e^{-i \vec k \vec r_\alpha}  
\end{align}
where we introduced the dynamical matrix $D_{il}(\vec k) = C_{ijkl} k_j k_k$. The resulting local strain field is given by 
\begin{align} \label{LocalStrain}
\varepsilon^{\rm loc}_{ij}(\vec r) =  g_{\Gamma_1}   
\int \frac{d\vec k}{(2\pi)^3} \mathcal{K}_{ij}(\vec k)
  \sum_\alpha e^{i \vec k (\vec r-\vec r_\alpha)} = g_{\Gamma_1}   \sum_\alpha  \tilde{\mathcal{K}}_{ij}(\vec r - \vec r_\alpha)
\end{align}
with the kernel $\mathcal{K}_{ij}(\vec k) = \frac{1}{2} (k_i  D^{-1}_{jn}(\vec k) k_n + k_j  D^{-1}_{in}(\vec k) k_n)$ and its Fourier transform $\tilde{\mathcal{K}}_{ij}$. It captures the long-range influence of the Pr impurities that is mediated by static strain fields.

Plugging this into Eq.~\eqref{QuadrupolarStaticStrain2} we then obtain for the macroscopic quadrupolar strain
\begin{align} \label{QuadrupolarStaticStrain}
\varepsilon_{u/v} = \frac{g_{\Gamma_1} g_{\Gamma_3}^2 \chi_Q}{(c^0_{11}-c^0_{12})/2} \frac{1}{V}\sum_{\alpha,\beta}
Q^{u/v}_{ij} \tilde{\mathcal{K}}_{ij}(\vec r_\alpha - \vec r_\beta)
= n_{\rm Pr}  \frac{ g_{\Gamma_1} g^2_{\Gamma_3}  \chi_Q}{(c^0_{11}-c^0_{12})/2} 
\int \frac{d\vec k}{(2\pi)^3}  Q^{u/v}_{ij} \mathcal{K}_{ij}(\vec k)  |z(\vec k) |^2.
\end{align}
where we abbreviated $z(\vec k) = \frac{1}{\sqrt{N_{\mathrm{Pr}}}}\sum_\alpha e^{i \vec k \vec r_\alpha}$ with $N_{\mathrm{Pr}}$ being the number of Pr ions. 

For wavevectors $|\vec k| L \gg 1$ much larger than the inverse linear system size and uniformly distributed positions of Pr ions, $z(\vec k)$ is a random complex number with a phase uniformly distributed on $[0,2\pi)$.
After disorder averaging $|z|^2 \to 1$, the above expression vanishes because the integral over wavevectors contains the quadrupolar matrix $Q^{u/v}_{ij}$. For a specific disorder configuration, however, the integral might be finite and for dimensional reasons it will scale as $1/a_{\mathrm{Pr}}^3$ where $a_{\mathrm{Pr}}$ is a characteristic length scale, that is expected to obey $a_{\mathrm{Pr}}^3 \sim 1/n_{\mathrm{Pr}}$. This suggests that the static local strains induced by Pr ions give rise to a  macroscopic quadrupolar strain $\varepsilon_{u/v} \sim n_{\mathrm{Pr}}^2 \chi_Q$ but with a prefactor suppressed by two powers of the Pr density. For this reasons, the local strains induced by the Pr ions themselves represents a sub-leading correction in the dilute limit.

 These disorder induced corrections, that break the cubic symmetry even in zero field, could in principle lead to systematic errors in our data analysis. In particular, care must be taken when identifying the bulk thermal expansion from uniaxial thermal expansion measurements along a single cubic direction $\langle 100 \rangle$ at zero field, see the discussion of Fig. 4 in the main text. Not only a small external stress from the sample holder but also the disorder might break the cubic symmetry inducing a finite quadrupolar macroscopic strain with an anomalous $T$-dependence induced by the response of the non-Kramers doublet. However, thermal expansion measurements along $\langle 111 \rangle$ will  not be affected by $\varepsilon_{u/v}$, see Eq.~\eqref{ref:Eq111}. Disorder might induce a finite shear, e.g. $\varepsilon_{xy}$, but without coupling to the non-Kramers doublet and, therefore, without the anomalous $T$-dependence due to the susceptibility $\chi_Q$.

\subsection{Anomalous bulk thermal expansion induced by quadrupolar Kondo correlations}

In the following subsections, we focus on the limit of both vanishing applied stress and magnetic fields. The linear coupling of the strain to the quadrupolar moment realizes an effective spin-boson model, and it is expected that this interaction eventually quenches the residual entropy of the 2CK fixed-point. There are two candidates for this quenching process:  the degeneracy of the non-Kramers doublet is either lifted by dynamical strain fields, i.e., by phonons, or by static local strains. Both lead to a perturbative correction to the free energy that resembles a high-temperature tail of a Schottky anomaly, however, here for the quench of the anomalous $\frac{k_B}{2} \log 2$ entropy. 

This process might be influenced by the application of a hydrostatic pressure. We will discuss possible thermodynamic signatures in the bulk thermal expansion and the specific heat. Whereas both quantities are expected to show an anomalous $T$-dependence, Gr\"uneisen scaling is actually expected. Possible reasons for a large Gr\"uneisen parameter is discussed.


\subsubsection{Renormalization of the elastic constant}

Integrating out the electronic quadrupolar degrees of freedom, one obtains a correction to the Hamiltonian in second order in the elastic coupling that reads for small strains
\begin{align} \label{EffHam}
\delta \mathcal{H}_{\rm strain} = -\frac{g^2_{\Gamma_3}\chi_Q }{2}   \sum_\alpha \Big( \varepsilon_u(\vec r_\alpha)^2 +  \varepsilon_v(\vec r_\alpha)^2\Big)
\end{align}
where we assumed that the quadrupolar moments of distinct Pr$^{3+}$ ions are uncorrelated. After disorder averaging, 
\begin{align} \label{EffHamAveraged}
\overline{\delta \mathcal{H}_{\rm strain}} = \int d\vec r  \Big(-\frac{g^2_{\Gamma_3} n_{\rm Pr} \chi_Q }{2} \Big( \varepsilon_u(\vec r)^2 +  \varepsilon_v(\vec r)^2\Big)\Big)
\end{align}
this correction reduces to a renormalization of the  elastic constant
\begin{align}
\frac{c_{11}-c_{12}}{2} = \frac{c^0_{11}-c^0_{12}}{2} - g^2_{\Gamma_3} n_{\rm Pr} \chi_Q.
\end{align}
The disorder averaging should be justified for strains that vary on a length scale that is much larger than the distance between Pr$^{3+}$ ions. As the susceptibility $\chi_Q \sim \log \frac{1}{T}$ diverges for low temperatures at the 2CK fixed-point, the renormalized elastic constant is reduced and might eventually vanishes at ultralow temperatures indicating the tendency towards a structural instability. The logarithmic 
$T$-dependence of this elastic constant has been confirmed by ultrasound experiments \cite{YanagisawaHidaka19}.

\subsubsection{Correction to the thermodynamics}

The effective Hamiltonian \eqref{EffHam} will lead to a correction to the free energy density
\begin{align}
\delta f = -\frac{1}{2} g^2_{\Gamma_3} n_{\rm Pr} \chi_Q  \langle \varepsilon^2_{E} \rangle_0
\end{align}
The finite expectation value $\langle \varepsilon^2_{E} \rangle_0$ might possess two different origins. First, it might arise from dynamic strain fields, i.e., phonon excitations. For long-wavelength phonons we can use the disorder averaged Hamiltonian \eqref{EffHamAveraged} 
and $\langle \varepsilon^2_{E} \rangle_0 \equiv \frac{1}{V} \int d\vec r \langle \varepsilon_u(\vec r)^2 +  \varepsilon_v(\vec r)^2 \rangle_0$. At lowest temperatures, this expectation value then corresponds to the zero-point fluctuations of phonons evaluated with the bare elastic constants. Second, it might arise from static local strain fields, $\langle \varepsilon^2_{E} \rangle_0 = \frac{1}{N_{\rm Pr}} \sum_\alpha \Big( \varepsilon_u(\vec r_\alpha)^2 +  \varepsilon_v(\vec r_\alpha)^2\Big)$ induced by crystal defects and inhomogeneities, where $N_{\rm Pr}$ is the number of Pr ions. In both cases, this yields a negative correction to the entropy 
\begin{align}
\delta S = - \partial_T \delta f = \frac{1}{2} g^2_{\Gamma_3} n_{\rm Pr}   \langle \varepsilon^2_{E} \rangle_0 \partial_T \chi_Q 
= - \frac{1}{2} g^2_{\Gamma_3} n_{\rm Pr}   \langle \varepsilon^2_{E} \rangle_0 A_Q \frac{1}{T}.
\end{align}
where we used  $\chi_Q \approx A_Q \log \frac{1}{T}$ with a constant $A_Q$. This should be compared to the  $-1/T^2$ behavior of the high-temperature tail of a standard Schottky anomaly. This entropy entails a singular correction to both the specific heat $\delta C = T \partial_T \delta S$ and the volume thermal expansion $\delta \beta = -\partial_p \delta S$ as a function of $T$,
\begin{align}
\delta C &= \frac{1}{2} g^2_{\Gamma_3} n_{\rm Pr} \langle \varepsilon^2_{E} \rangle_0 A_Q   \frac{1}{T},\qquad
\delta \beta = 
\frac{1}{2} \Big[\partial_p  \Big(g^2_{\Gamma_3} n_{\rm Pr} \langle \varepsilon^2_{E} \rangle_0  A_Q \Big) \Big] \frac{1}{T}
\end{align}
where $\partial_p$ is the derivative with respect to hydrostatic pressure. The ratio of both yield a reduced  Gr\"uneisen parameter 
\begin{align}
\delta \Gamma = \frac{\delta \beta}{\delta C} = \partial_p \log (g^2_{\Gamma_3} n_{\rm Pr} \langle \varepsilon^2_{E} \rangle_0 A_Q)
\end{align} 
that is determined by the pressure dependence of the product of characteristic parameters $g^2_{\Gamma_3} n_{\rm Pr} \langle \varepsilon^2_{E} \rangle_0 A_Q$. In case that, e.g., the main $p$-dependence arises from the elastic coupling $g_{\Gamma_3}$, the reduced Gr\"uneisen parameter reduces to $\delta \Gamma =  \partial_p \log g^2_{\Gamma_3}$.

\section{Analysis of the experimental data}

\subsection{Thermal expansion and Grüneisen parameter for $\boldsymbol{H\parallel [001]}$}

This section discusses the longitudinal and transverse thermal expansion coefficients that were measured for $\boldsymbol{H}\parallel[001]$. Based on these two measurements, the volume and quadrupolar thermal expansion coefficients can be deduced. In addition, the specific heat data, used for the calculation of the quadrupolar Grüneisen ratio, is presented. 
\begin{figure}[h]
\includegraphics[width=0.43\textwidth]{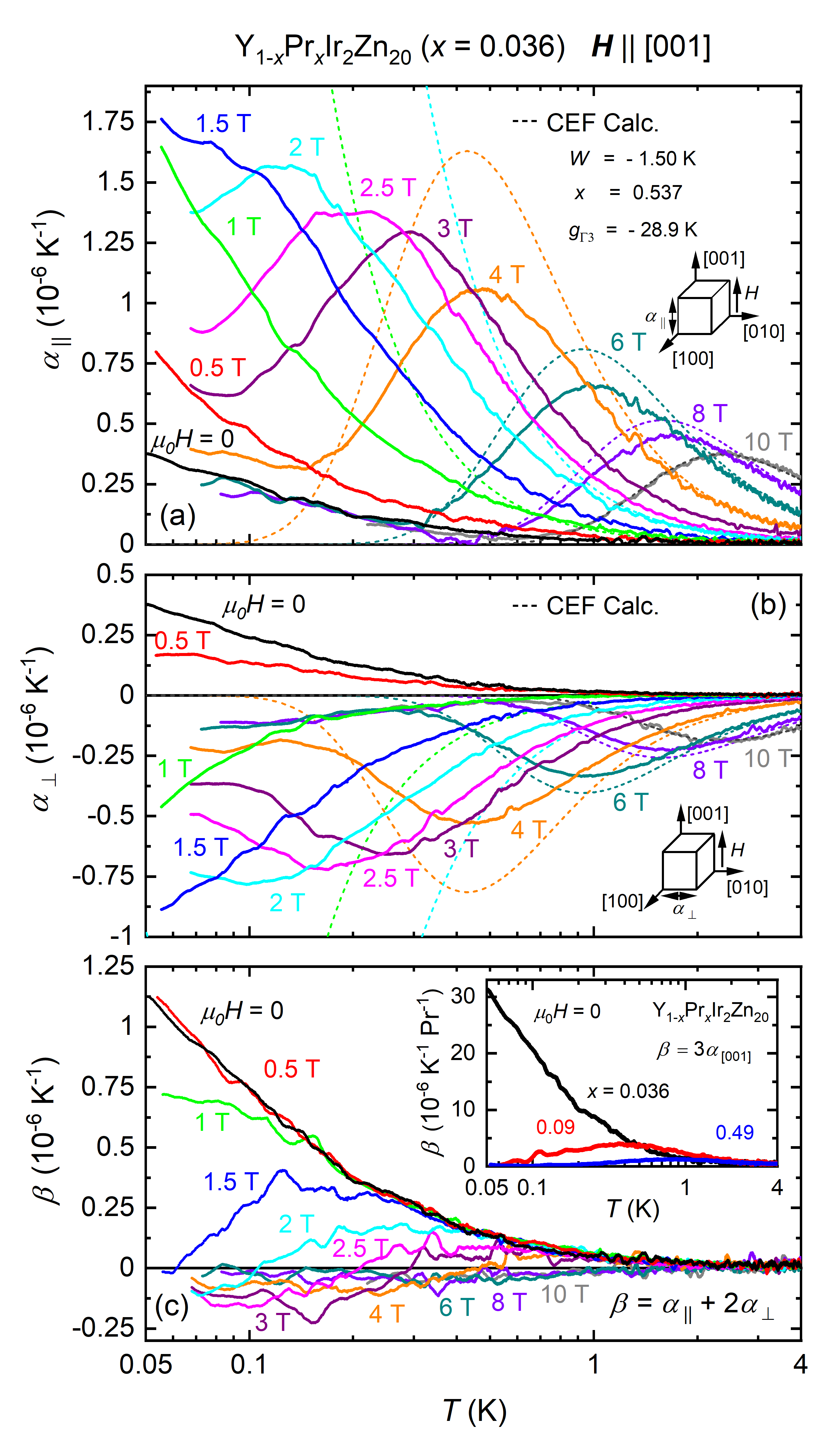}
\caption{Temperature dependence of the (a) longitudinal thermal expansion coefficient $\alpha_{\parallel}$ and (b) transverse thermal expansion coefficient $\alpha_{\perp}$  together with CEF calculations for $\boldsymbol{H} \parallel [001]$. (c) Temperature dependence of the volume thermal expansion coefficient $\beta=\alpha_{\parallel}+2\alpha_{\perp}$ at different magnetic fields. The inset shows the temperature dependence of $\beta=3\alpha_{[001]}$ normalized to the Pr-concentration $x$ for three differently doped samples at zero magnetic field.} 
\label{fig:figS2}
\end{figure}
\subsubsection{Longitudinal, transverse, volume and quadrupolar thermal expansion}
In order to deduce the quadrupolar and volume thermal expansion coefficients $\alpha_{\mathrm{u}}$ and $\beta$, which are shown in Fig. 1(a) and Fig. 4(a) of the main text, longitudinal $(\alpha_{\parallel}=\alpha_{[001]} \parallel H)$ and transverse $(\alpha_{\perp}=\alpha_{[100], [010]} \parallel H)$ thermal expansion measurements were carried out for $\boldsymbol{H}\parallel [001]$. 

Figure \ref{fig:figS2}(a) and (b) shows the longitudinal ($\alpha_{\parallel}$) and transverse thermal expansion coefficients ($\alpha_{\perp}$) at various magnetic fields together with CEF simulations, which were calculated by following the approach explicitly discussed in section II. The quadrupole-strain coupling constant $g_{\Gamma_3}$ was determined at -28.9\,K by fitting the simulated data to the high field data at 10\,T where the system is expected to be in the fully localized state. This value is comparable to $g_{\mathrm{\Gamma_3}}=-38.0$\,K \cite{WoerlOnimaru19} of pure PrIr$_2$Zn$_{20}$.

By using the longitudinal and transverse thermal expansion coefficients, the volume thermal expansion coefficient, as shown in Fig. \ref{fig:figS2}(c), is obtained
\begin{equation}
\beta=\alpha_{\parallel}+2\alpha_{\perp}.
\end{equation} 
By using $\alpha_{\parallel}$ and $\beta$, the quadrupolar thermal expansion coefficient $\alpha_{\mathrm{u}}$ then calculates with the help of Eq. (\ref{ref:Eq5}) as
\begin{equation}
\alpha_{\mathrm{u}}= \sqrt{3}\left(\alpha_{\parallel}-\frac{1}{3}\beta \right),
\end{equation}
which is presented in Fig. 1(a) of the main text. 




\subsubsection{Grüneisen Parameter}
\begin{figure}[h]
\includegraphics[width=0.48\textwidth]{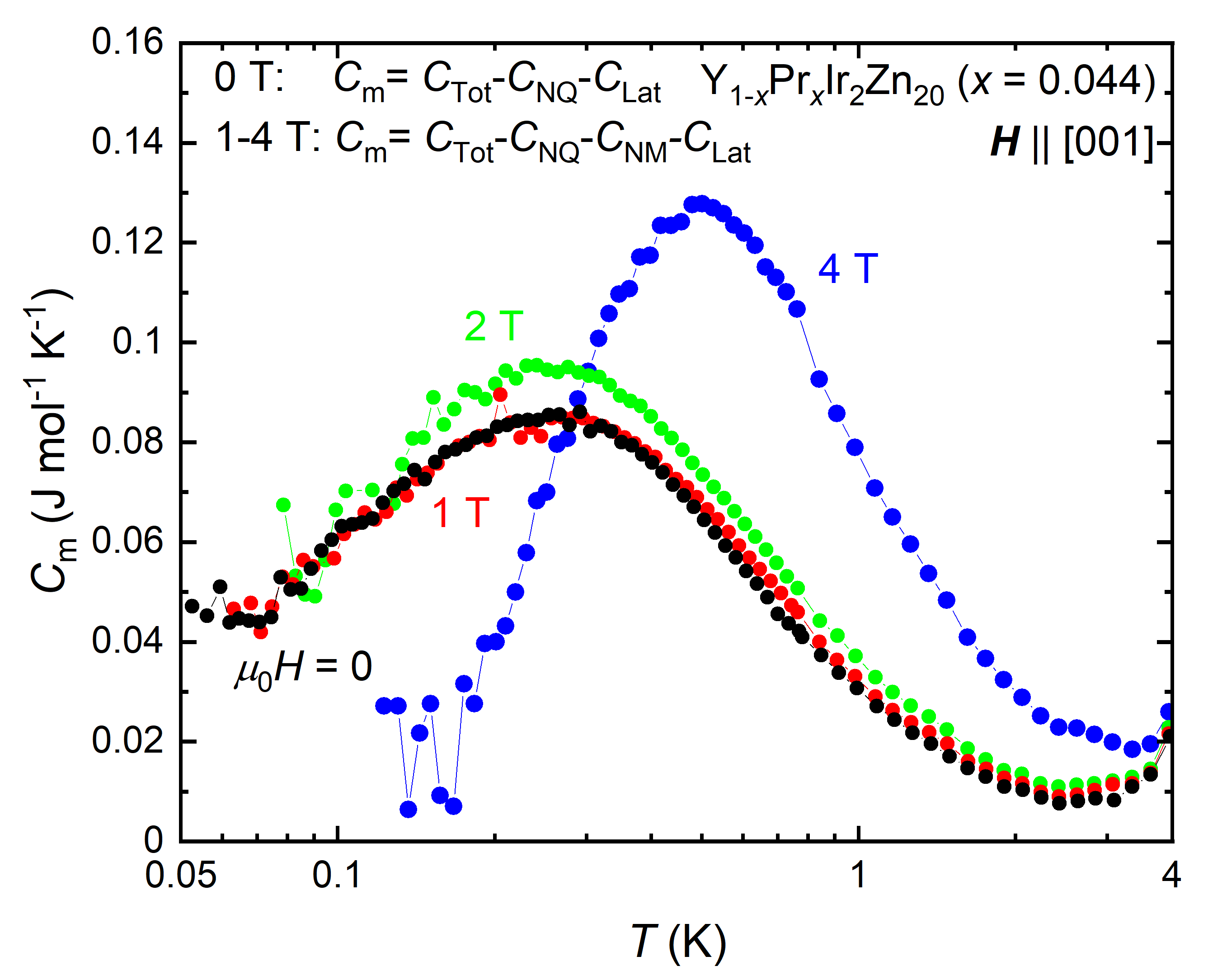}
\caption{Temperature dependence of the 4f contribution to the molar specific heat $C_{\mathrm{m}}$ at magnetic fields $H\parallel[001]$ up to 4\,T. The specific heat of this particular single crystal has already been shown in Ref. \cite{YamaneOnimaru18, YamaneOnimaru18AIP} (except data at 1\,T). The here presented data was partially taken from Refs. \cite{YamaneOnimaru18, YamaneOnimaru18AIP} ($T\ge0.8$\,K) and partially remeasured ($T<0.8$\,K).}
\label{fig:figS3}
\end{figure}

In order to calculated the quadrupolar Grüneisen parameter plotted in Fig. 1(b) of the main text, the specific heat data measured on a Y$_{1-x}$Pr$_x$Ir$_2$Zn$_{20}$ single crystal from the same batch as the one used for the thermal expansion measurement, having a comparable Pr-concentration of $x=0.044$, was used. The specific heat of this particular sample has already been published in Refs.~\cite{YamaneOnimaru18, YamaneOnimaru18AIP}. Subsequently shown data was partially taken from Refs.~\cite{YamaneOnimaru18, YamaneOnimaru18AIP} ($T\gtrsim0.8)$ and partially remeasured with a different setup to obtain higher resolved data down to lower temperatures ($T<0.8$\,K). Figure \ref{fig:figS3} shows the 4f contribution to the molar specific heat as a function of temperature of Y$_{1-x}$Pr$_x$Ir$_2$Zn$_{20}$ ($x=0.044$) at various magnetic fields  $\boldsymbol{H}\parallel [001]$.

In zero magnetic field a quadrupolar nuclear contribution and a lattice contribution, determined from a reference measurement on YIr$_2$Zn$_{20}$, were subtracted. For the data obtained in magnetic field, a magnetic nuclear Schottky contribution was subtracted additionally. As outlined previously by Yamane et al. \cite{YamaneOnimaru18}, this particular sample shows a behavior $C/T \sim \log 1/T$ in zero magnetic field consistent with the single-impurity quadrupolar Kondo scenario.

The molar volume, used for the calculation of the Grüneisen parameter, was determined by using the specific heat sample's lattice constant $a=14.197\times10^{-10}$\,m and the number of formula units per unit volume $Z=8$, as 
\begin{equation}
V_{\mathrm{m}}(x=0.044)=\frac{V_{\mathrm{m}}\times a^3}{Z}=2.154024\times10^{-4}\,\frac{\mathrm{m}^3}{\mathrm{mol}}
\end{equation}

As no specific heat measurement was performed for $\mu_0 H=0.5$\,T, the specific heat data at zero magnetic field was used to calculate the quadrupolar Grüneisen parameter at 0.5\,T. Fig. \ref{fig:figS3} shows that the specific heat at 0\,T and 1\,T are nearly identical, which suggests that $C_{\mathrm{m}}$ can be considered as nearly field independent for $\mu_0 H\le 1$\,T. This justifies the use of the zero field specific heat for the calculation of $\Gamma_{\mathrm{u}}$ at 0.5\,T. For all other $\Gamma_{\mathrm{u}}$ curves shown in Fig. 1(b) in  the main text, thermal expansion and specific heat were measured at the same magnetic field.  As pointed out in Section I.A., by taking the error in the Pr content of the highly diluted samples into account, the Pr content of the two single crystals employed for the dilatometry ($x=0.033$ and $x=0.036$) can be considered as the same as the Pr content of the single crystal employed for the specific heat measurement ($x=0.044$). Consequently, a normalization of the thermal expansion and the specific heat to the Pr content was not necessary in order to calculate the Grüneisen parameter. 

\subsection{Magnetostriction for $\boldsymbol{H\parallel[001]}$}
In this section we present the longitudinal and transverse magnetostriction coefficients measured for $\boldsymbol{H}\parallel[001]$ on highly diluted Y$_{1-x}$Pr$_x$Ir$_2$Zn$_{20}$ $(x=0.036)$. From these measurements, the volume magnetostriction and the quadrupolar magnetostriction can be determined. Finally, we discuss how the low temperature magnetostriction can be employed to estimate the quadrupole susceptibility.
\begin{figure}[h]
       \centering
       \includegraphics[width=0.45\textwidth]{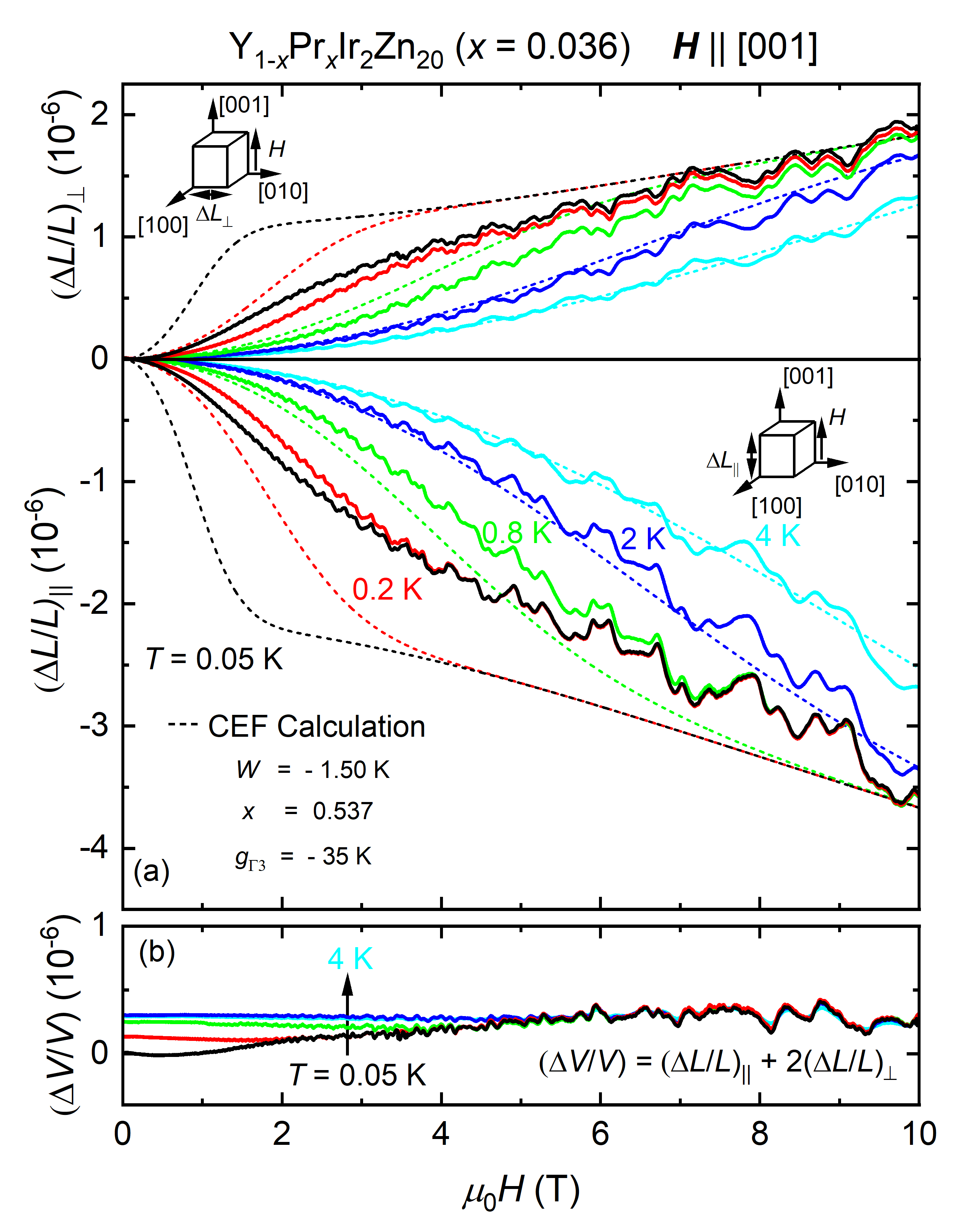}
       \caption{(a) Longitudinal and transverse magnetostriction coefficients for $\boldsymbol{H}\parallel[001]$, measured on the highly diluted Y$_{1-x}$Pr$_x$Ir$_2$Zn$_{20}$ single crystal with $x=0.036$. (b) Volume magnetostriction determined from the longitudinal and transverse magnetostriction shown in (a) via the relation $\Delta V/V=\Delta L/L_{\parallel}+2\Delta L/L_{\perp}$.}
       \label{fig:figS4}
 \end{figure}
 
\subsubsection{Longitudinal, transverse, volume and quadrupolar magnetostriction}
To determine the volume and quadrupolar magnetostriction coefficients $\varepsilon_{\mathrm{B}}$ and $\varepsilon_{\mathrm{u}}$ of the the highly diluted Y$_{1-x}$Pr$_x$Ir$_2$Zn$_{20}$, longitudinal ($(\Delta L/L)_{\parallel}=\varepsilon_{\parallel}=\varepsilon_{[001]} \parallel H$) and transverse ($(\Delta L/L)_{\perp}=\varepsilon_{\perp}=\varepsilon_{[100],[010]}\perp H$) magnetostriction measurements were performed which are shown in Fig. \ref{fig:figS4}. CEF calculation are shown as dashed lines. The curves were calculated by use of the same CEF parameter $W=-1.50$\,K and $x=0.537$ by which also the thermal expansion was simulated. For the background elastic modulus the value reported for highly diluted Y$_{1-x}$Pr$_x$Ir$_2$Zn$_{20}$ in the Supplemental material of Ref.\,\cite{YanagisawaHidaka19}, namely $(C_{11}-C_{12})/2=52.771$\,GPa. The magnetostriction analysis suggests a slightly larger quadrupole strain coupling constant of $g_{\Gamma_3}=-35.0$\,K than the thermal expansion ($g_{\Gamma_3}=-28.9$\,K). The Pr density per unit volume is $N_{\mathrm{Pr}}(x=0.036)=0.101\times10^{27}$\,m$^{-3}$. $\varepsilon_{\parallel}$ and $\varepsilon_{\perp}$ allow for the 
calculation of the bulk strain $\varepsilon_{\mathrm{B}}=\Delta V/V$ via the relation
\begin{equation}
\varepsilon_{\mathrm{B}}=\varepsilon_{\parallel}+2\varepsilon_{\perp}.
\end{equation}
The bulk strain  determined in this manner is shown in Fig. \ref{fig:figS4}(b) as a function of magnetic field $\boldsymbol{H}\parallel[001]$ for various temperatures. As compared to the longitudinal and transverse magnetostriction, the volume change is markedly small, which is in line with the previous results on pure PrIr$_2$Zn$_{20}$ \cite{WoerlOnimaru19}.  At the lowest measured temperatures, $(\Delta V/V)$ exhibits a small magnetic field dependence which is gradually suppressed as temperature increases. Thereby, $(\Delta V/V)$ changes the most at intermediate field ranging from 1\,T to 3\,T, which can be likely connected to the suppression of the divergent volume thermal expansion in magnetic field.  The field variation of $(\Delta V/V)$ at different temperatures is similar to the one of the previously reported ($c_{11}-c_{12}$)/2 elastic constant \cite{YanagisawaHidaka19}, providing further indication for a possible coupling between volume strain and the quadrupole susceptibility via static or dynamic strain fields.

The quadrupolar magnetostriction $\varepsilon_{\mathrm{u}}$, which is shown in Fig. 3 of the main text, can simply be calculated by using Eq.~\eqref{ref:Eq5} as
\begin{equation}
\varepsilon_{\mathrm{u}}=\sqrt{3}\left(\varepsilon_{\parallel}-\frac{\varepsilon_{\mathrm{B}}}{3}\right).
\end{equation}

\subsubsection{Quadrupole susceptibility}

Next, we briefly discuss the estimation of the quadrupole susceptibility from the magnetostriction data. As outlined in the previous section, the expectation value of the quadrupole operator  $\left\langle O^0_2 \right\rangle$ depends for small $H$ quadratically on magnetic field
\begin{equation}
\left\langle O^0_2 \right\rangle=\chi^F_{\mathrm{Q}}(\mu_{0}H)^2,
\label{ref:Eq9}
\end{equation}
where $\chi^F_Q \sim \chi_Q$. This initial quadratic field dependence is thereby mainly generated by the mixing of $\Gamma_3$ ground state and the first excited $\Gamma_4$ state in magnetic field. Plugging this into Eq.~\eqref{ref:Eq4} one obtains 
\begin{equation}
\varepsilon_{\mathrm{u}}=\frac{N_{\mathrm{Pr}}{g_{\Gamma_3}}}{(c^0_{11}-c^0_{12})/2} \chi_{\mathrm{Q}}(\mu_{0}H)^2.
\label{ref:Eq10}
\end{equation}

\begin{figure}[t]
\includegraphics[width=0.48\textwidth]{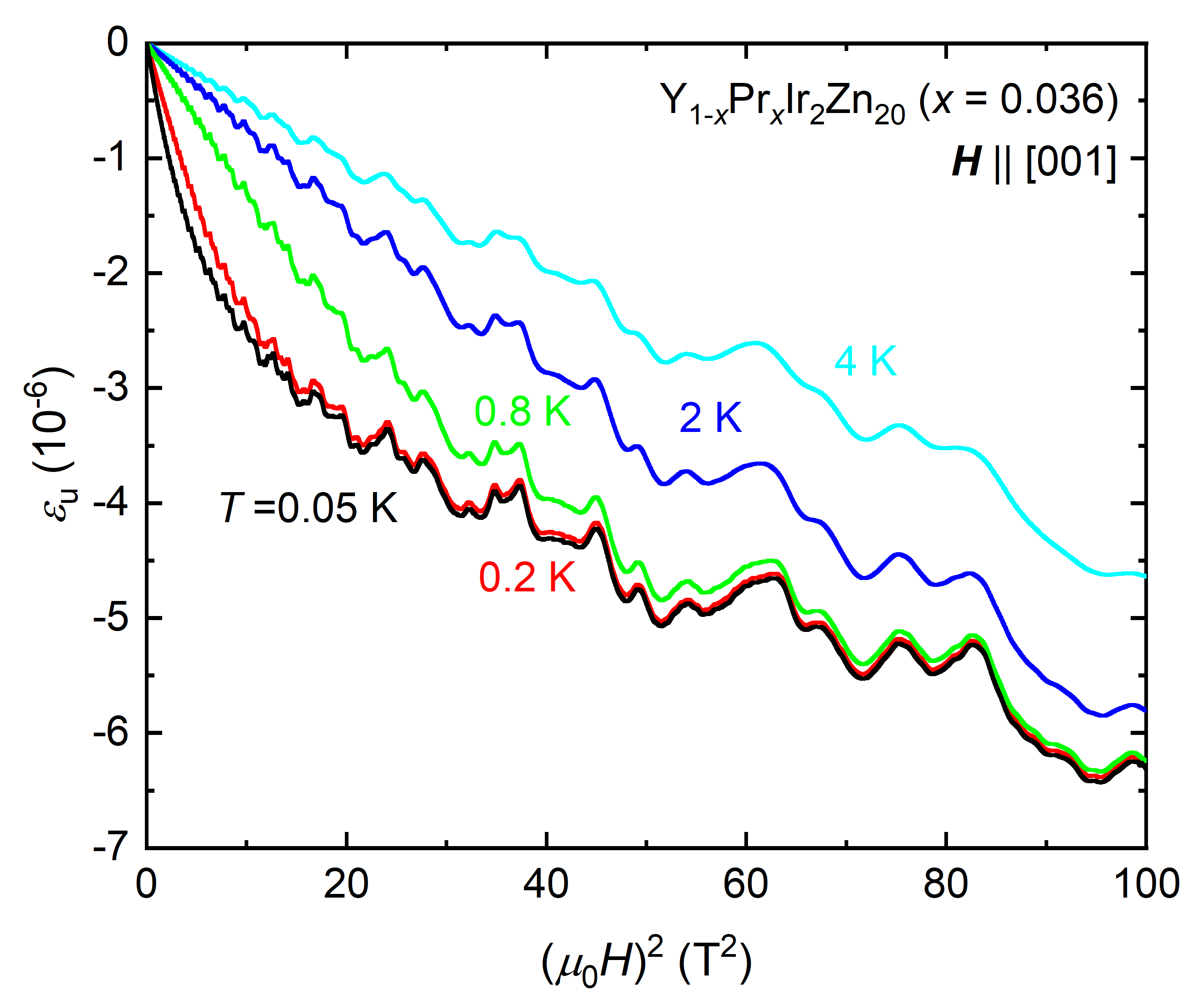}
\caption{$\varepsilon_{\mathrm{u}}$ as a function of $(\mu_0 H)^2$ at different temperatures to illustrate the quadratic field dependence of $\epsilon_{\mathrm{u}}$ for small $\boldsymbol{H}\parallel[001]$.}
\label{fig:figS5}
\end{figure}

This suggests that the initial quadratic field dependence of $\varepsilon_{\mathrm{u}}$ measured at various temperatures, provides direct access to the quadrupole susceptibility $\chi_{\mathrm{Q}}$. 
Fig. \ref{fig:figS5} shows $\varepsilon_{\mathrm{u}}$ versus $(\mu_{0}H)^2$.  
By fitting the low field magnetostriction with a linear regression that yields the initial slope $m$, we can then determine $\chi_{\mathrm{Q}}$ at selected temperatures. For $\chi_{\mathrm{Q}}$ it follows
\begin{equation}
\chi_{\mathrm{Q}}=\frac{(c_{11}^{0}-c_{12}^{0})/2}{N_{\mathrm{Pr}}g_{\Gamma_3}}m
\end{equation}

In the inset of Fig. 3 of the main text, we also present $\chi_{\mathrm{Q}}$ of two higher doped Y$_{1-x}$Pr$_{x}$Ir$_2$Zn$_{20}$ single crystals with $x=0.09$ and $x=0.49$. As for these two samples the magnetostriction was only measured parallel to magnetic field,  $\chi_{\mathrm{Q}}$ is determined by evaluating the strain parallel to magnetic field 
\begin{equation}
\varepsilon_{\parallel}=\frac{1}{\sqrt{3}}\frac{N_{\mathrm{Pr}}{g_{\Gamma_3}}}{(C^0_{11}-C^0_{12})/2}\chi_{\mathrm{Q}}^{\mathrm{F}}H^2,
\label{ref:Eq13}
\end{equation}
under the assumption that $\varepsilon_{\mathrm{v}}=0$ and that the field dependent contribution of the isotropic bulk strain $\varepsilon_{\mathrm{B}}$ is small as compared to the uniaxial quadrupolar contribution  $\varepsilon_{\mathrm{u}}$. The former assumption is justified, as $\epsilon_{\mathrm{u}} \gg \epsilon_{\mathrm{B}}$ for both the highly diluted system with $x=0.036$ and the pure system PrIr$_2$Zn$_{20}$ \cite{WoerlOnimaru19}.

\end{document}